\documentclass [twocolumn,pra,showkeys,nofootinbib]{revtex4-1} 
\usepackage[english]{babel}
\usepackage[utf8]{inputenc}
\usepackage{graphicx}
\usepackage{comment}
\usepackage{hyperref}
\usepackage{multirow}
\usepackage{color}
\usepackage{amsmath}
\usepackage{array}
\usepackage{subfig}
\usepackage{tikz}
\usetikzlibrary{arrows,shapes}

\input{Qcircuit} 

\begin{document}

\title{Adiabatic state preparation study of methylene}
\author{Libor Veis}
\email{libor.veis@jh-inst.cas.cz}
\affiliation{J. Heyrovsk\'{y} Institute of Physical Chemistry, Academy of Sciences of the Czech \mbox{Republic, v.v.i.}, Dolej\v{s}kova 3, 18223 Prague 8, Czech Republic}

\author{Ji\v{r}\'{i} Pittner}
\email{jiri.pittner@jh-inst.cas.cz}
\affiliation{J. Heyrovsk\'{y} Institute of Physical Chemistry, Academy of Sciences of the Czech \mbox{Republic, v.v.i.}, Dolej\v{s}kova 3, 18223 Prague 8, Czech Republic}

\date{\today}

\begin{abstract}
Quantum computers attract much attention as they promise to outperform their classical counterparts in solving certain type of problems. One of them with practical applications in quantum chemistry is simulation of complex quantum systems. An essential ingredient of \textit{efficient} quantum simulation algorithms are initial guesses of the exact wave functions with high enough fidelity. As was proposed in [Aspuru-Guzik et al., \textit{Science} \textbf{309}, 1704 (2005)], the exact ground states can in principle be prepared by the adiabatic state preparation method. Here, we apply this approach to preparation of the lowest lying multireference singlet electronic state of methylene and numerically investigate preparation of this state at different molecular geometries. We then propose modifications that lead to speeding up the preparation process. Finally, we decompose the minimal adiabatic state preparation employing the direct mapping in terms of two-qubit interactions.

\end{abstract}


\maketitle

\section{Introduction}

Quantum computers are appealing for their ability to solve certain type of problems more effectively than in the classical setting \cite{shor_1994, grover_1997}. A prominent example is the integer factorization \cite{shor_1994} where they offer an exponential speedup, which has far-reaching consequences for cryptography. The original idea of a quantum computer indeed belongs to R. Feynman and Y. Manin \cite{feynman_1982, manin_1980}, who pointed out that quantum computers could in principle be used for the \textit{efficient} (polynomially scaling) simulation of complex quantum systems \cite{lloyd_1996, zalka_1998, ortiz_2001, somma_2002, abrams_1997, abrams_1999, ovrum_2007}. This idea, which has practical applications also in quantum chemistry, employs mapping of the Hilbert space of a studied quantum system onto the Hilbert space of a register of quantum bits (qubits), both of them being exponentially large.

The past few years have witnessed a remarkable interest in the application of quantum computing for solving problems in quantum chemistry and the list of relevant papers is already quite rich. We will mention only some of them and for a complete list refer the reader to recent reviews \cite{kassal_review, yung_review, veis_2012_chapter}.

The first work connecting quantum computation and quantum chemistry concerned the \textit{efficient} calculations of thermal rate constants \cite{lidar_1999}. Aspuru-Guzik et al. in their seminal paper \cite{aspuru-guzik_2005} presented the \textit{efficient} quantum algorithm for ground state molecular energy calculations, a quantum analogue of the classical full configuration interaction (FCI) method. They also proposed preparation of the exact ground states for such computations by the adiabatic evolution, a method which we investigate in this paper. Since these two pioneering works, other theoretical papers involving e.g. calculations of excited states \cite{wang_2008}, quantum chemical dynamics \cite{kassal_2008}, calculations of molecular properties \cite{kassal_2009}, or calculations of relativistic systems \cite{veis_2012} were published. Application of adiabatic quantum computing for finding low energy conformations of proteins was described in \cite{perdomo_2008, perdomo-ortiz_2012}.

As was mentioned in \cite{wecker_2013}, quantum computers with about 100 (noise free) qubits would exceed the limits of classical full configuration interaction (FCI) calculations dramatically. This is in contrast to other quantum algorithms, e.g. the Shor's algorithm \cite{shor_1994, shor_1997} for integer factorization would for practical tasks in cryptography require thousands of qubits. For this reason, calculations and simulations of quantum systems will belong to the first practical applications of quantum computers, which is also supported by recent proof-of-principle few-qubit experiments \cite{lanyon_2010, du_2010, li_2011, lu_2011, lanyon_2011, peruzzo_2013}. Several improvements reducing the resource requirements of fault-tolerant implementation and thus paving the way for practical simulations were presented in \cite{jones_2012}. Universal programmable circuit schemes for emulating an operator with the application to hydrogen molecule computation were introduced in \cite{daskin_2012}.

The quantum FCI algorithm \cite{aspuru-guzik_2005, whitfield_2010} requires an initial guess of the exact eigenstate, whose quality influences the success probability of measuring the desired energy. This can be either a classical approximation [e.g. complete active space (CAS) based wave function \cite{wang_2008, veis_2010}], an exact state prepared by the adiabatic state preparation method \cite{aspuru-guzik_2005}, which we investigate here, or by the algorithmic cooling method \cite{xu_2012}, or also an unitary coupled cluster approximation optimized by the recently presented combined classical-quantum variational approach \cite{peruzzo_2013, yung_2013}. 

In this paper, we are dealing only with a preparation of the exact ground states. The desired information like energy can be obtained either with the standard phase estimation algorithm (PEA) \cite{nielsen_chuang} or other methods developed to reduce qubit and coherence time requirements \cite{lanyon_2010, biamonte_2011, li_2011, peruzzo_2013}, i.e. to adapt the procedure for a present-day or near-future quantum technology. 

The structure of this paper is as follows: First, we briefly introduce the adiabatic quantum computing and adiabatic state preparation method in Section \ref{aqc}. In Section \ref{ch2}, we present our model system - methylene (CH$_2$), the results of classical simulations of the lowest lying singlet electronic state preparations, and suggest modifications that speed up the preparation process. Section \ref{experimental_proposal} involves analysis of the minimal CH$_2$ adiabatic state preparation employing the direct mapping when using two-body qubit Hamiltonians.

\section{Adiabatic quantum computing}
\label{aqc}

The original idea of a quantum computation by adiabatic evolution, usually denoted as adiabatic quantum computing (AQC), is due to Farhi et al. \cite{fahri_2000, fahri_2000b, fahri_science_2001}. In AQC, one slowly varies the Hamiltonian of a quantum register starting with a simple one ($H_{\mathrm{init}}$), whose ground state is easy to prepare, and ending with the final one ($H_{\mathrm{final}}$), whose ground state encodes the solution to the problem. If the gap between the two lowest energy levels is large enough along the path and the change is slow enough, the register has a tendency to remain in its ground state according to the quantum adiabatic theorem \cite{messiah}.

To be more precise, a quantum state of $n$ qubits $\ket{\psi(t)}$ evolves in time according to the Schr\"odinger equation

\begin{equation}
  i \hbar \frac{d}{dt} \ket{\psi(t)} = H(t) \ket{\psi(t)},
  \label{schrodinger_equation}
\end{equation}

\noindent
where $H(t)$ is the time-dependent Hamiltonian operator which equals $H_{\mathrm{init}}$ for $t = 0$ and $H_{\mathrm{final}}$ for $t = T$ and $T$ is a total time of the process.

One of the possible adiabatic evolution paths is a linear interpolation

\begin{equation}
  H(s) = (1 - s) H_{\mathrm{init}} + s H_{\mathrm{final}},   
  \label{ham}
\end{equation}

\noindent
where $s = t/T$. 

A sufficient condition for the total time to ensure that $\ket{\psi(t)}$ follows the ground state of $H(t)$ adiabatically in this case reads

\begin{equation}
  T \gg \frac{\epsilon}{g_{\mathrm{min}}^2},
  \label{asp_time}
\end{equation}

\noindent
with the minimum energy gap, $g_{\mathrm{min}}$, defined by

\begin{equation}
  g_{\mathrm{min}} = \min_{0 \le s \le 1} [E_{1}(s) - E_{0}(s)],
  \label{gmin}
\end{equation}

\noindent
and $\epsilon$ in the form

\begin{equation}
  \epsilon = \max_{0 \le s \le 1} \Big| \bra{l=1; s} \frac{d H}{ds} \ket{l=0; s} \Big|,
  \label{epsilon}
\end{equation}

\noindent
where states $\ket{l=0/1; s}$ denotes ground/first excited state of $H(s)$.

AQC is a model of quantum computation that is alternative to the most common approach, the quantum circuit model \cite{nielsen_chuang}. Both of them are equivalent in terms of computational power, which means that one can simulate the other with only a polynomial overhead \cite{Aharonov, kempe_2006, mizel_2007}. But as AQC is based on ground states, it is expected to be naturally more robust against noise and may offer to perform medium-size simulations \cite{aspuru_guzik_review} without a need of sophisticated methods of quantum error correction. 

The only experimentally programmable qubit interactions are two-body. Therefore, when implementing AQC experimentally, one has to transform the Hamiltonian (\ref{ham}) of a quantum register to contain at most 2-qubit interactions. One of the possibilities of transforming general Hamiltonians containing non-commuting $k$-qubit terms to 2-qubit terms are methods of perturbative gadgets \cite{kempe_2006, biamonte_2008, jordan_2008, cao_2013}. We will use this approach in Section \ref{experimental_proposal} to transform 4-qubit terms to 2-qubit terms in case of the small CH$_2$ experimental proposal.

If one has not the direct access to the Hamiltonian of a quantum register, AQC can still be efficiently simulated on a digital quantum computer as the time evolution with the time dependent Hamiltonian (\ref{ham}). When the Hamiltonian contains non-commuting terms, the Trotter-Suzuki approximations \cite{trotter} have to be employed. Furthermore, in this case the dependence of the total time of the adiabatic evolution on the energy gap can be improved from $g_{\mathrm{min}}^{-2}$ to $g_{\mathrm{min}}^{-1}$, which is also optimal \cite{boixo_2010}.

\subsection{Adiabatic state preparation}

When using AQC for the purpose of initial state preparation for subsequent quantum simulation, we speak about the adiabatic state preparation (ASP). This method was proposed by Aspuru-Guzik et al. in their seminal paper \cite{aspuru-guzik_2005} as a method for preparation of the exact ground states of molecular Hamiltonians for energy computations by the phase estimation algorithm (PEA). It is also an essential part of the quantum algorithm employing adiabatic non-destructive measurements of energy (or other constants of motion) \cite{biamonte_2011}.

Before discussing the original ASP procedure \cite{aspuru-guzik_2005}, we have to say few words about the mapping of a quantum chemical wave function onto a quantum register. The most convenient and simply scalable approach is so-called direct mapping \cite{aspuru-guzik_2005}. In this case, individual spin orbitals (or Kramers pair bispinors in a relativistic generalization \cite{veis_2012}) are directly assigned to qubits, because each spin orbital can be either occupied or unoccupied, corresponding to $\ket{1}$ or $\ket{0}$ states. Motivated by the need to employ as few qubits as possible in the first experimental realizations, compact mappings from a subspace of fixed-electron-number wave functions, spin-adapted \cite{aspuru-guzik_2005} or symmetry-adapted \cite{wang_2008, veis_2012} wave functions to the register of qubits have also been proposed.

The final ASP molecular Hamiltonians can be \textit{efficiently} expressed in the second-quantized form \cite{szabo_ostlund} as

\begin{equation}
  H_{\mathrm{final}} = \sum_{pq} h_{pq} a^{\dagger}_{p} a_{q} + \frac{1}{2}\sum_{pqrs} \langle pq | rs \rangle a^{\dagger}_{p} a^{\dagger}_{q} a_{s} a_{r},
  \label{ham_sec_quant}
\end{equation}

\noindent
where $h_{pq}$ and $\langle pq | rs \rangle$ are one- and two-electron integrals (in the ``physicist's notation" \cite{szabo_ostlund}) in the molecular spin orbital basis. Generally, when using the direct mapping, the Jordan-Wigner \cite{whitfield_2010,jordan_1928} or the Bravyi-Kitaev \cite{bravyi_2002,seeley_2012} transformations can be used to map the fermionic creation and annihilation operators to spin operators represented by the Pauli $\sigma$-matrices. Such mappings correctly preserve the fermionic anti-commutation relations. In fact, when implementing ASP directly, i.e. when transforming the Hamiltonian to contain at most 2-qubit interactions, the Bravyi-Kitaev approach is crucial for the \textit{efficiency} of the algorithm \cite{babbush_2013}.

In \cite{aspuru-guzik_2005}, the initial ASP Hamiltonians were represented by matrices with all matrix elements equal to zero, except $\mathbf{H}_{11}$, which was equal to the Hartree-Fock energy ($E_{\mathrm{HF}}$). Expressed in the direct mapping, the initial Hamiltonians are represented by matrices with all matrix elements equal to zero, except $\mathbf{H}_{ii} = E_{\mathrm{HF}}$, where $i$ corresponds to the computational basis state that represents the Hartree-Fock Slater determinant. This single basis state, which is easy to prepare, is also the initial state of the ASP algorithm.

For reasons that will be discussed further, we propose another type of initial ASP Hamiltonians, namely the Hamiltonians equal to a sum of the Fock operators \cite{szabo_ostlund}. In the canonical restricted Hartree-Fock (RHF) spin orbital basis, they have the following diagonal form

\begin{eqnarray}
  H_{\mathrm{init}, \mathrm{MP}} & = & \sum_{p} f_{pp} a^{\dagger}_{p} a_{p} , \label{h_mp} \\
  f_{pp} & = & h_{pp} + \sum_{i \in \text{occ.}} \Big( \langle p i | p i \rangle - \langle p i | i p \rangle \Big). \\ \nonumber
\end{eqnarray}

\noindent
Such Hamiltonians are taken as unperturbed in the M{\o}ller-Plesset type of the Rayleigh-Schr\"odinger perturbation theory \cite{moller_plesset,szabo_ostlund} (therefore the abbreviation MP). The ground states of $H_{\mathrm{init}, \mathrm{MP}}$ in the subspace matching the correct number of electrons are again the single computational basis states representing the Hartree-Fock Slater determinants.

At the end of this section, we have to note that finding the exact ground state energy of a general two-body Hamiltonian is known to be QMA-complete \cite{kempe_2006}, i.e it is supposed to be difficult even for a quantum computer. The complexity of this task manifests itself in the dependence of the total ASP time on the energy gap between the ground and excited states (Eq. \ref{asp_time}) and as far as the gap is not sufficiently large, these states cannot be prepared \textit{efficiently}. In fact there are physical systems in nature, such as spin glasses, that may never fall down into their ground states. On the other hand, most molecules can on physical grounds be expected to have large enough energy gaps \cite{biamonte_2011} and their ground states thus should be preparable by the ASP method. 

Apart from preparation of the exact ground states, ASP can in principle be used also to improve the ground state fidelity of the initial guess wave functions for subsequent PEA computations \cite{oh_2008}.

\section{Simulated ASP of $\tilde{a}~^{1}A_{1}$ state of CH$_{2}$}
\label{ch2}

We have numerically studied ASP of the $\tilde{a}~^{1}A_{1}$ state of methylene. Despite not being the true ground state of (\ref{ham_sec_quant}) at the equilibrium geometry (it is $\tilde{X}~^{3}B_{1}$), this state is the lowest singlet electronic state and thus can be prepared by ASP when starting with a singlet initial state (e.g. the closed shell Hartree-Fock Slater determinant). Note that this assumption is valid only in the non-relativistic regime, where the molecular Hamiltonian (\ref{ham_sec_quant}) commutes with the square of the total spin operator.

\begin{figure}
\begin{tabular}{ m{1.5cm} m{1cm} m{4.2cm} }
\begin{tikzpicture}[scale=0.5]
  \draw [thick] (0,0) to (1.6,0);
  \draw [->,thick,shorten <=3pt,shorten >=3pt] (0.6,-0.5) -- (0.6,0.5);
  \draw [->,thick,shorten <=3pt,shorten >=3pt] (1.0,0.5) -- (1.0,-0.5);
  \draw [thick] (0,1) to (1.6,1);
  \draw [->,thick,shorten <=3pt,shorten >=3pt] (0.6,0.5) -- (0.6,1.5);
  \draw [->,thick,shorten <=3pt,shorten >=3pt] (1.0,1.5) -- (1.0,0.5);
  \draw [thick] (0,2) to (1.6,2);
  \draw [->,thick,shorten <=3pt,shorten >=3pt] (0.6,1.5) -- (0.6,2.5);
  \draw [->,thick,shorten <=3pt,shorten >=3pt] (1.0,2.5) -- (1.0,1.5);
  \draw [thick,red] (0,3) to (1.6,3);
  \draw [->,thick,shorten <=3pt,shorten >=3pt,red] (0.6,2.5) -- (0.6,3.5);
  \draw [->,thick,shorten <=3pt,shorten >=3pt,red] (1.0,3.5) -- (1.0,2.5);
  \draw [thick,red] (0,4) to (1.6,4);
  \node at (2.5,0) {$1a_1$};
  \node at (2.5,1) {$2a_1$};
  \node at (2.5,2) {$1b_2$};
  \node [red] at (2.5,3) {$3a_1$};
  \node [red] at (2.5,4) {$1b_1$};
  \node at (0.9,-1.0) {$\ket{\phi_1}$};
\end{tikzpicture}
&
\begin{tikzpicture}[scale=0.5]
  \draw [thick] (0,0) to (1.6,0);
  \draw [->,thick,shorten <=3pt,shorten >=3pt] (0.6,-0.5) -- (0.6,0.5);
  \draw [->,thick,shorten <=3pt,shorten >=3pt] (1.0,0.5) -- (1.0,-0.5);
  \draw [thick] (0,1) to (1.6,1);
  \draw [->,thick,shorten <=3pt,shorten >=3pt] (0.6,0.5) -- (0.6,1.5);
  \draw [->,thick,shorten <=3pt,shorten >=3pt] (1.0,1.5) -- (1.0,0.5);
  \draw [thick] (0,2) to (1.6,2);
  \draw [->,thick,shorten <=3pt,shorten >=3pt] (0.6,1.5) -- (0.6,2.5);
  \draw [->,thick,shorten <=3pt,shorten >=3pt] (1.0,2.5) -- (1.0,1.5);
  \draw [thick,red] (0,3) to (1.6,3);
  \draw [thick,red] (0,4) to (1.6,4);
  \draw [->,thick,shorten <=3pt,shorten >=3pt,red] (0.6,3.5) -- (0.6,4.5);
  \draw [->,thick,shorten <=3pt,shorten >=3pt,red] (1.0,4.5) -- (1.0,3.5);
  \node at (0.9,-1.0) {$\ket{\phi_2}$};
\end{tikzpicture}
& $\begin{array}{ccc}\ket{\psi_{\text{bent}}} & \sim & 0.95 \ket{\phi_1} - 0.15 \ket{\phi_2} \\ \ket{\psi_{\text{linear}}} & \sim & \color{red} 0.7 \color{black}  \ket{\phi_1} - \color{red} 0.7 \color{black} \ket{\phi_2} \end{array} $ 
\end{tabular}
\caption{Dominant electronic configurations $\ket{\phi_1}$ and $\ket{\phi_2}$ of the $\tilde{a}~^{1}A_{1}$ state wave function. $\ket{\psi_{\text{bent}}}$ corresponds to the equilibrium geometry, whereas $\ket{\psi_{\text{linear}}}$ to the linear one, where $3a_1$ and $1b_1$ orbitals become degenerate.}
\label{el_configurations}
\end{figure}
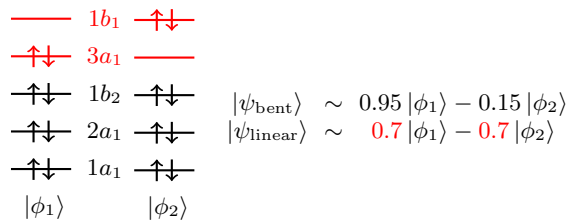

\begin{figure*}
  \subfloat[][]{
  \hskip -1cm
    \includegraphics[width=8.5cm]{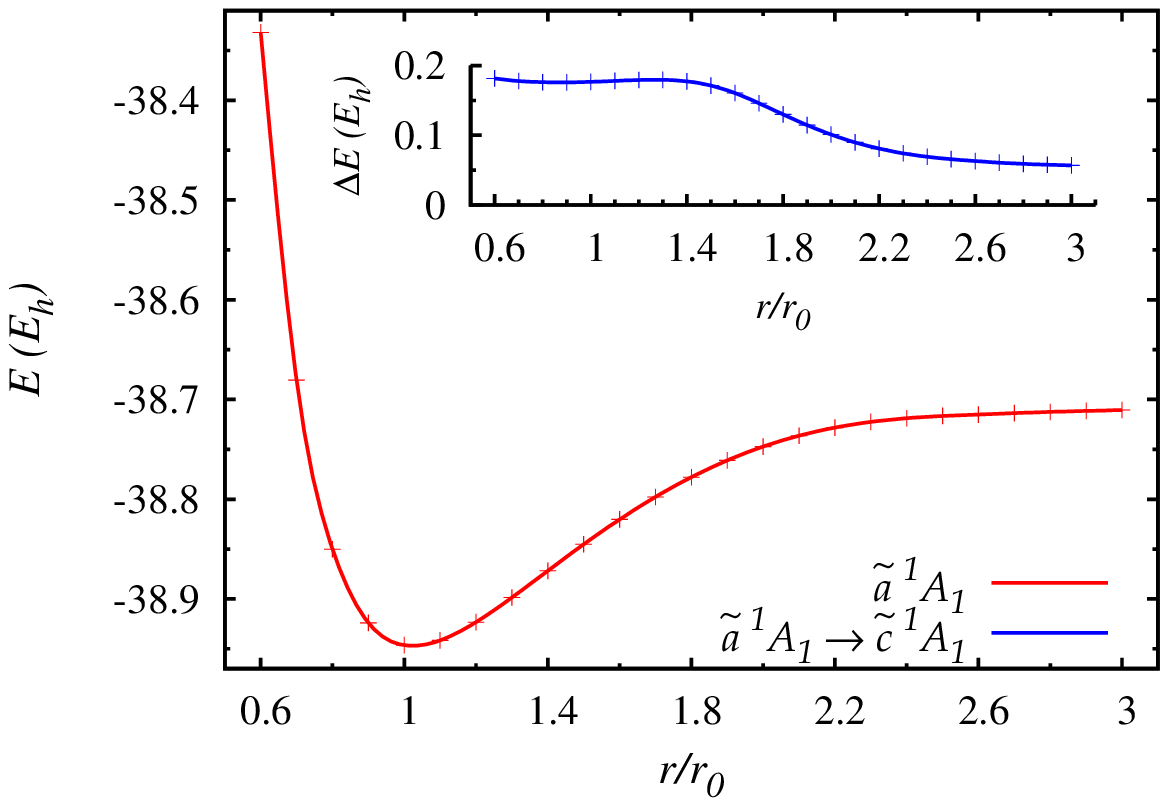}
    \label{en_disociace}
  }
  \subfloat[][]{
    \includegraphics[width=8.5cm]{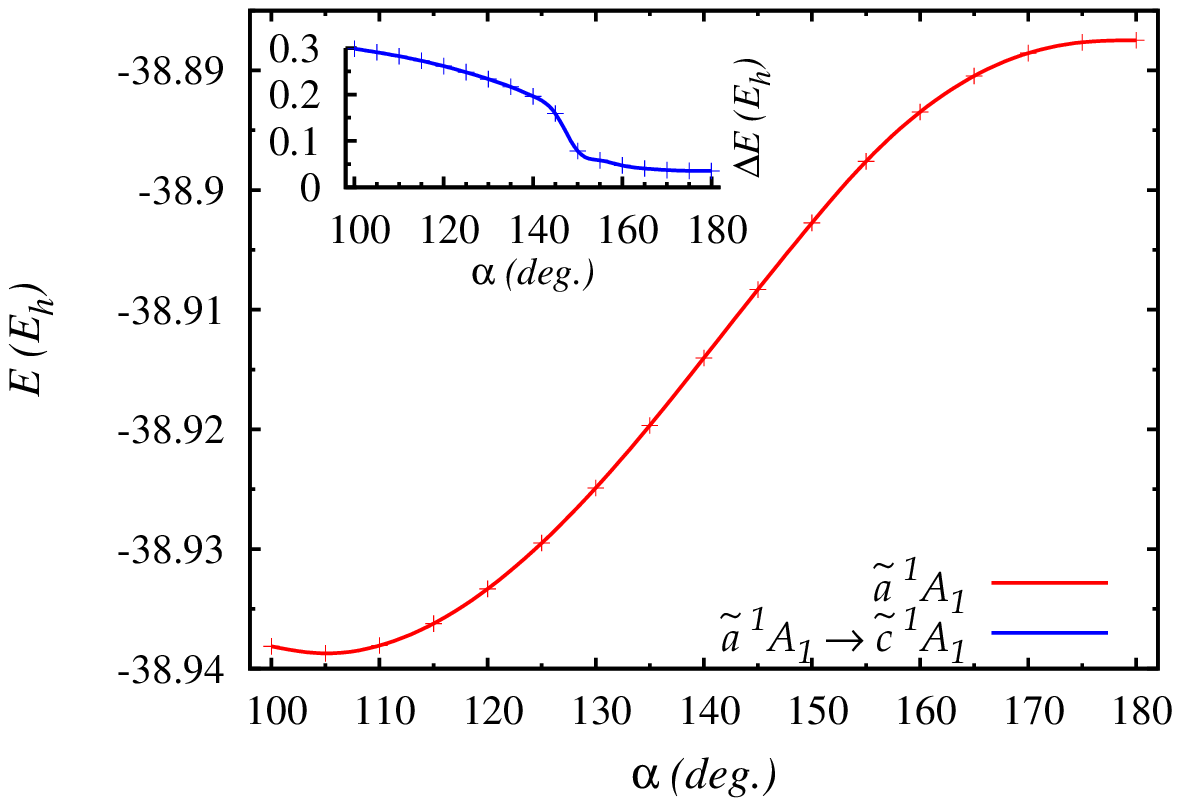}
    \label{en_bending}
  }
  \caption{Energies of the lowest-lying singlet electronic state ($\tilde{a}^{1}A_{1}$) and energy gap between $\tilde{a}^{1}A_{1}$ and $\tilde{c}^{1}A_{1}$ states for (a) C-H bond stretching [CASCI(6,12)/cc-pVTZ] and (b) H-C-H angle bending [CASCI(6,11)/cc-pVTZ]. $r_0$ denotes the equilibrium bond distance and $\alpha$ the H-C-H angle.}
  \label{en_graphs}
\end{figure*}

The $\tilde{a}~^{1}A_{1}$ state of methylene is well known for its multireference character, which makes it a suitable benchmark system for testing of newly developed computational methods (see, e.g. \cite{bwcc1, evangelista-allen1, mkcc_our2, demel-pittner-bwccsdt}). We are in fact following our previous paper \cite{veis_2010}, where we used CH$_2$ as a benchmark for simulations of computations with the quantum FCI method based on the iterative phase estimation algorithm (IPEA).

Figure \ref{el_configurations} shows the dominant electronic configurations that contribute to the $\tilde{a}~^{1}A_{1}$ state wave function. The multireference character of this state is a consequence of the quasi-degeneracy of the boundary orbitals $3a_1$ and $1b_1$. As in \cite{veis_2010}, we have simulated two processes: C-H bond stretching and H-C-H angle bending. These processes were chosen designedly because correct description of bond breaking is generally a difficult task and H-C-H angle bending due to the very strong multireference character at linear geometries (see Figure \ref{el_configurations}).

Energies and energy gaps between the two lowest singlet electronic states calculated at the same level of theory at which the ASP simulations has been carried out (see Section \ref{computational_details} below) for both simulated processes are shown in Figure \ref{en_graphs}. It can be seen that when going to more stretched C-H bonds or linear geometries, the energy gaps between the lowest singlet states decreases. One may therefore expect that these regions of the potential energy surface will be for ASP more problematic. 

Our aim was to numerically investigate these regions and suggest modifications of the original ASP procedure \cite{aspuru-guzik_2005} that would lead to decreasing of the total ASP time. As will be discussed further, it turns out that MP-type of initial Hamiltonians (\ref{h_mp}) are advantageous for this purpose. We have also tested initial states based on small-CAS-like wave functions that cover the major part of a static correlation. Such states can be on a quantum computer prepared \textit{efficiently} \cite{wang_2009}. Last but not least, we have investigated non-linear interpolation paths which draw on the specific knowledge of how the ground state changes during ASP.

\subsection{Computational details}
\label{computational_details}

\begin{figure*}
  \subfloat[][]{
  \hskip -1cm
    \includegraphics[width=8.5cm]{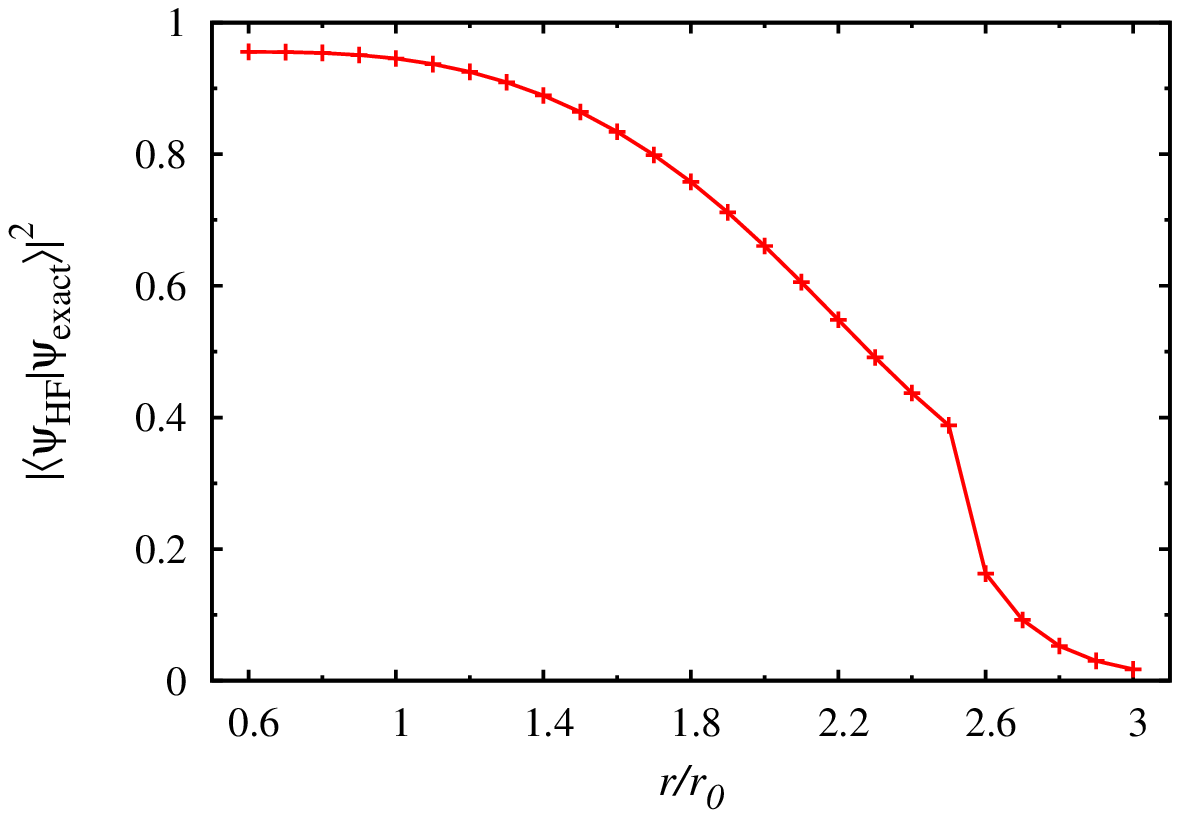}
    \label{overlap_disociace}
  }
  \subfloat[][]{
    \includegraphics[width=8.5cm]{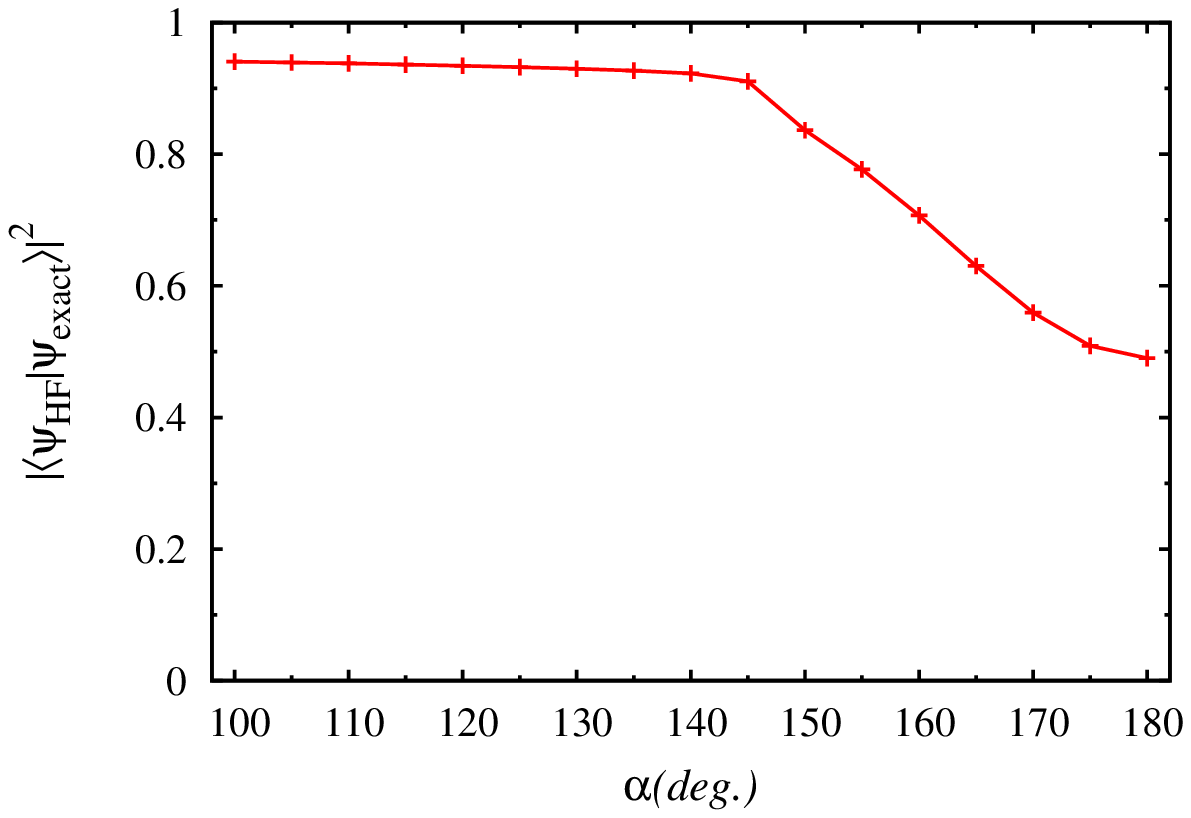}
    \label{overlap_bending}
  }
  \caption{Squared overlap between the HF initial and the exact wave functions (for definition, see the text) for (a) C-H bond stretching and (b) H-C-H angle bending. $r_0$ denotes the equilibrium bond distance and $\alpha$ the H-C-H angle.}
  \label{overlaps}
\end{figure*}

In all our simulations, we used the cc-pVTZ basis set \cite{dunning_1989}. We, of course, could not manage to simulate ASP with FCI Hamiltonians (\ref{ham_sec_quant}) in such a large basis. To keep the size of the problem still tractable, we modelled the FCI by the complete active space configuration interaction (CASCI) with a limited CAS. For computational reasons, we employed the compact mapping using the spin and point-group symmetries. 

In case of the C-H bond stretching, we were using the CASCI with 6 electrons in 12 orbitals [CAS(6,12)]. This leads to roughly 4300 configuration state functions (CSFs) in the aforementioned compact mapping. For H-C-H angle bending, we had to use 6 electrons in just 11 orbitals [CAS(6,11)], because CAS(6,12) would at the linear geometry include only one of the two degenerate orbitals. 

Regarding simulations of ASP on a classical computer, we basically numerically integrated the time-dependent Schr\"odinger equation (\ref{schrodinger_equation}) in the basis of CASCI CSFs with the time-dependent Hamiltonian (\ref{ham}). We therefore numerically propagated the wave function according to

\begin{equation}
  \ket{\Psi(t+\Delta t)} = e^{-i H(t + \frac{\Delta t}{2}) \Delta t} \ket{\Psi(t)},
\end{equation}

\noindent
where for the action of an exponential of a Hamiltonian on a vector representing the wave function of a quantum register, we employed the \textsc{Dalton} program's \cite{dalton,dalton2} direct CI routine performing the CASCI Hamiltonian matrix vector multiplications. The energy scale was therefore shifted by the core and nuclear repulsion contributions. 

For technical reasons, we restricted ourselves to CASCI with 6 electrons in 7 orbitals when using the MP-type of initial Hamiltonians (\ref{h_mp}). In these simulations, we employed less compact mapping from the subspace of wave functions with constant number of alpha ($N_{\text{alpha}}$) and beta ($N_{\text{beta}}$) electrons, $N_{\text{alpha}} = N_{\text{beta}}$. 

To verify the validity of our conclusions concerning the comparison of both types of initial Hamiltonians, we also simulated the original ASP \cite{aspuru-guzik_2005} in CAS(6,7) space and checked that the results in both orbital spaces do not differ.

In order to get a clue of how well ASP will perform for systems in which we have only a very poor initial guess of the exact wave function, we also numerically simulated ASP with initial Hamiltonian of the following form \cite{perdomo_2008}

\begin{equation}
  H_{\text{init,X}} = \sum_{i}^{n_{\text{qubits}}} \frac{1}{2} \big( I - \sigma_x^i \big),
	\label{hlx}
\end{equation}

\noindent
where $\sigma_x^i$ denotes Pauli $x$ matrix acting on $i$th qubit. Such a Hamiltonian has a non-degenerate ground state equal to the homogeneous superposition of all the computational basis states and thus does not presume any information about the true wave function. Moreover, it is trivial to prepare this state experimentally. As in case of MP-type of initial Hamiltonians, FCI was modelled by CASCI(6,7).

The equilibrium geometry of CH$_2$ was adopted from \cite{sherrill_1997} and corresponded to $r_{e} = 1.1089~\rm{\AA}$ and $\alpha_{e} = 101.89^{~\circ}$). 

\subsection{Results}
\label{results}

The squared overlaps between the HF initial wave functions and the exact wave functions modelled by the CAS(6,12) for the bond stretching and CAS(6,11) for the angle bending processes, which influence the length of the ASP procedure are presented in Figure \ref{overlaps}.


\begin{figure*}
  \subfloat[][]{
  \hskip -1cm
    \includegraphics[width=8.5cm]{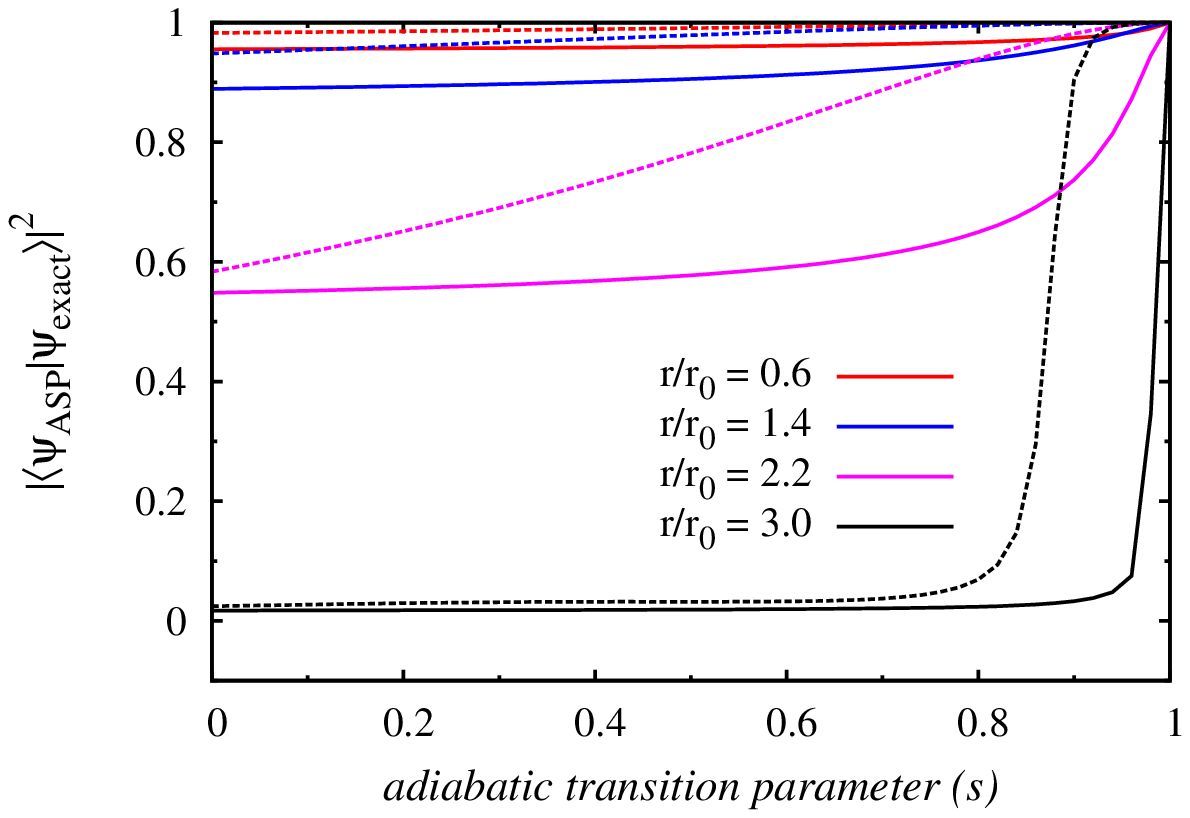}
    \label{asp_disociace}
  }
  \subfloat[][]{
    \includegraphics[width=8.5cm]{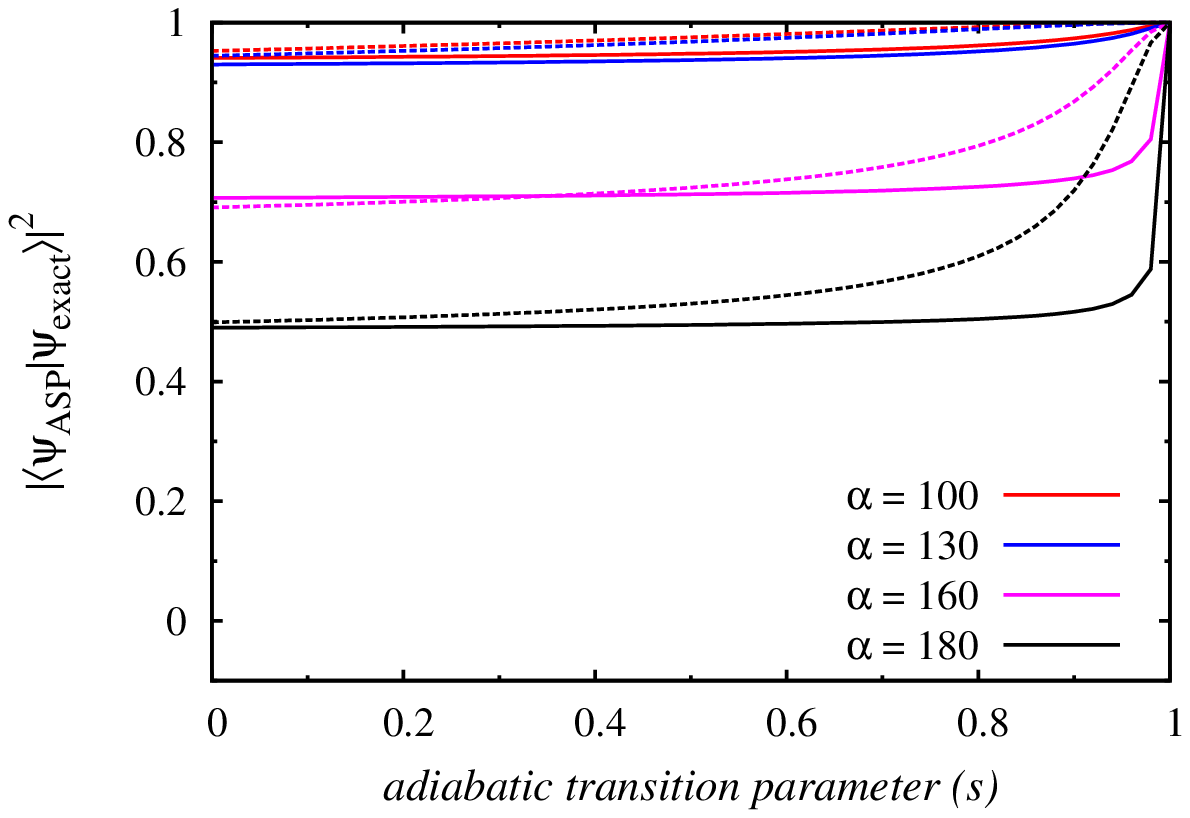}
    \label{asp_bending}
  }
  \caption{Squared overlap between the adiabatically prepared and the exact wave functions (for definition, see the text) for (a) C-H bond stretching and (b) H-C-H angle bending. Solid lines correspond to the original Aspuru-Guzik type of initial Hamiltonians \cite{aspuru-guzik_2005}, dashed lines correspond to the M{\o}ller-Plesset type of initial Hamiltonians (\ref{h_mp}).}
  \label{asp_graphs}
\end{figure*}

\begin{figure*}
  \subfloat[][]{
  \hskip -1cm
    \includegraphics[width=8.5cm]{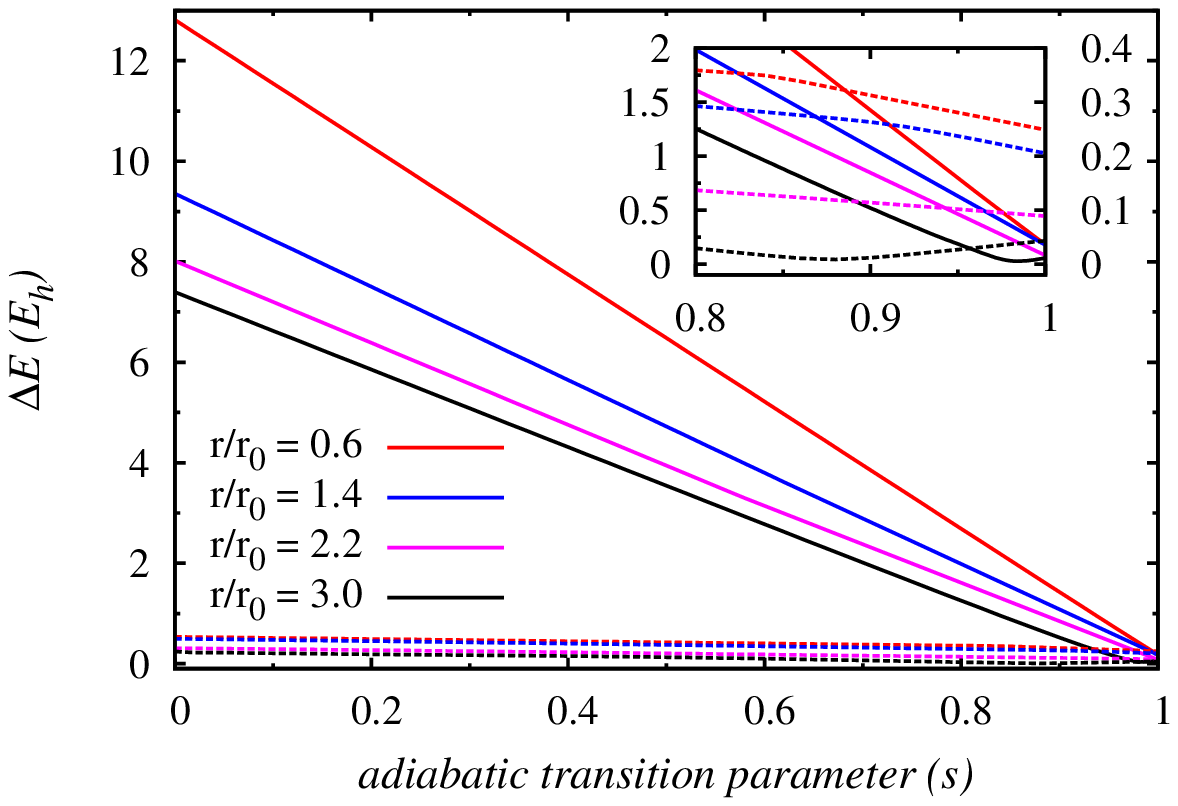}
    \label{en_gap_disociace}
  }
  \subfloat[][]{
    \includegraphics[width=8.5cm]{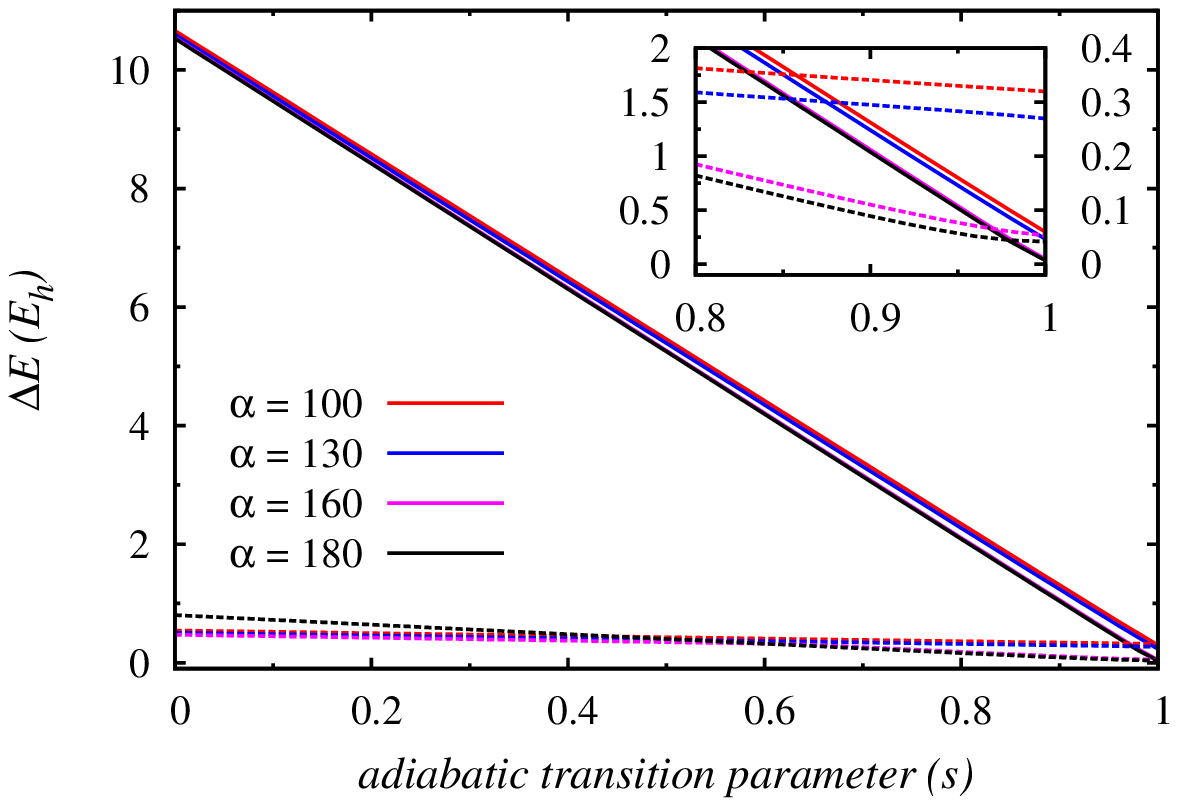}
    \label{en_gap_bending}
  }
  \caption{Dependence of the energy gaps $\Delta E$ between $\tilde{a}^{1}A_{1}$ and $\tilde{c}^{1}A_{1}$ states of CH$_2$ on the adiabatic transition parameter $s$ (for the level of theory employed, see the text) for (a) C-H bond stretching and (b) H-C-H angle bending. Solid lines correspond to the original Aspuru-Guzik type of initial Hamiltonians \cite{aspuru-guzik_2005}, dashed lines correspond to the M{\o}ller-Plesset type of initial Hamiltonians (\ref{h_mp}). The left energy scale of insets belong to solid lines whereas the right to dotted lines.} 
  \label{en_gap_graphs}
\end{figure*}

In Figure \ref{asp_graphs}, the dependence of the squared overlap between the adiabatically prepared wave functions and the exact wave functions on the adiabatic transition parameter $s$ is shown for the representative geometries of the two simulated processes. Both types of initial Hamiltonians are compared. The exact wave functions correspond to the CAS(6,12)/CAS(6,11) in case of the original initial Hamiltonians \cite{aspuru-guzik_2005} and to the CAS(6,7) in case of the MP-type of initial Hamiltonians (\ref{h_mp}).

Figure \ref{en_gap_graphs} presents the change of the energy gaps between the two lowest singlet $A_{1}$ states ($\tilde{a}^{1}A_{1} \rightarrow \tilde{c}^{1}A_{1}$) during the ASP procedure. The dependence of the energy gaps on the adiabatic transition parameter $s$ is depicted for the same geometries as in Figure \ref{asp_graphs}. Also both types of initial Hamiltonians are tested. It can be seen that the minimal energy gaps are independent of a type of the initial Hamiltonian and correspond to $s = 1$.

\begin{figure}
  \includegraphics[width=8.5cm]{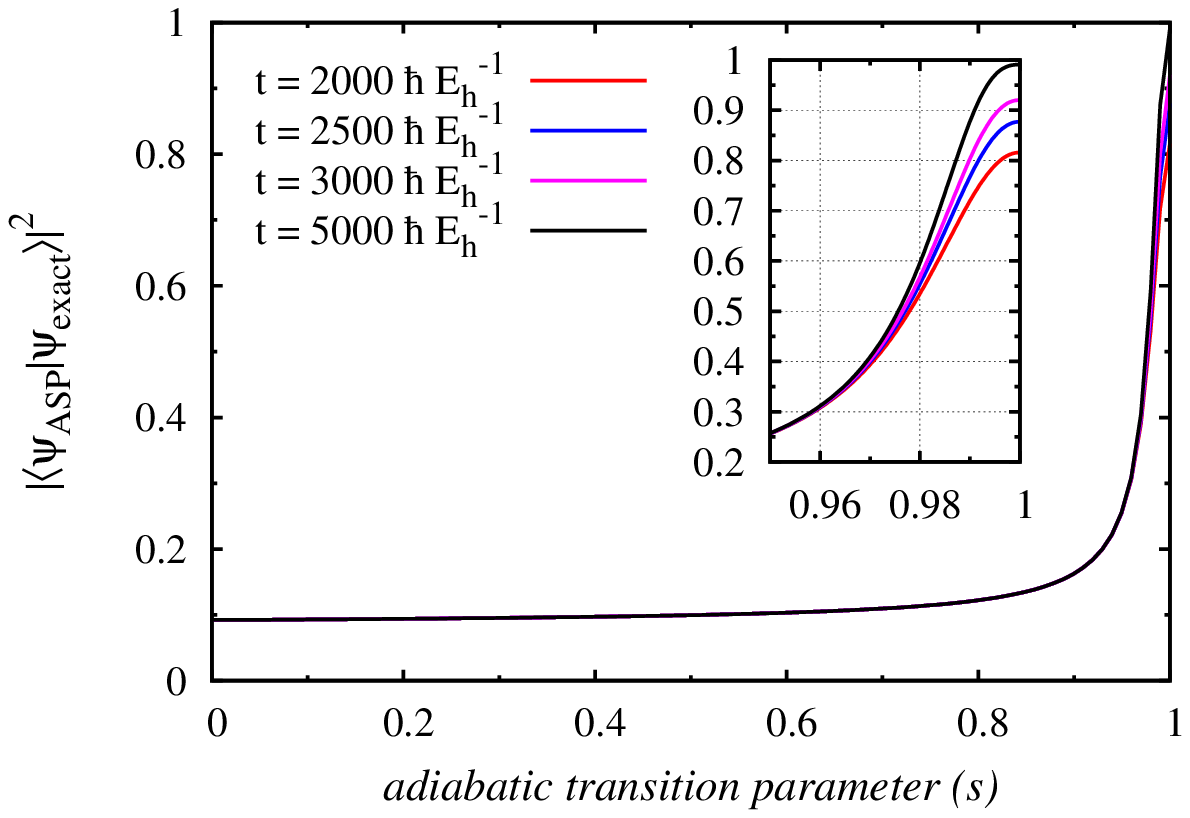}
  \caption{Dependence of the squared overlap between the ASP wave function and the exact wave function [CAS(6,12)] for C-H bond stretching with $r/r_0 = 2.7$ on the adiabatic transition parameter $s$. Different lengths of ASP with total time expressed in atomic units ($\hbar E_{h}^{-1}$) are presented.}
  \label{time_convergence}
\end{figure}

Figure \ref{time_convergence} shows the dependence of the squared overlap between the adiabatically prepared wave functions and the exact [CAS(6,12)] wave functions on the adiabatic transition parameter $s$ for different total times of the ASP procedure. The presented results correspond to the C-H bond stretching geometry with $r/r_0 = 2.7$. One can see that the most problematic part, which is difficult to follow adiabatically, is the steep change at the end of the process. The time is expressed in atomic units [$1~a.u.~(\hbar E_{h}^{-1}) \approx 10^{-17}~s$].

\begin{figure*}
  \subfloat[][]{
    \hskip -1cm
    \includegraphics[width=8.5cm]{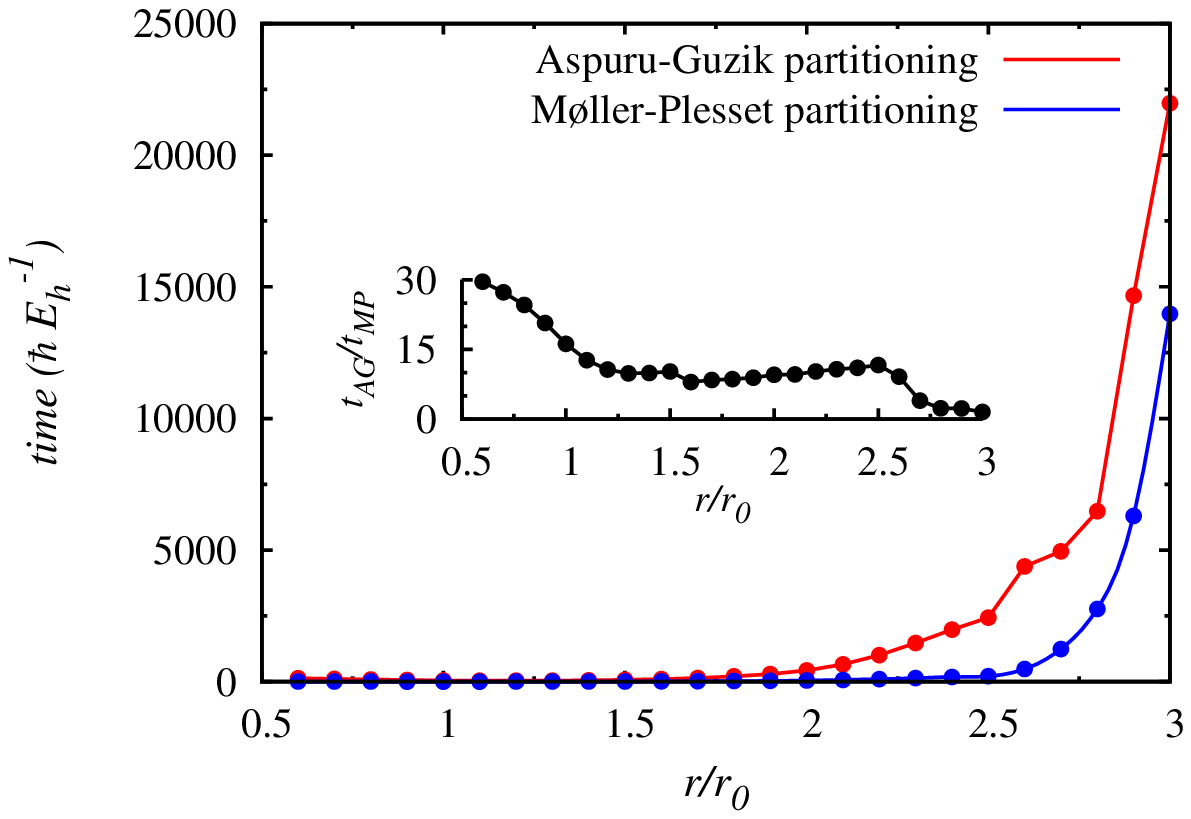}
    \label{time_disociace}
  }
  \subfloat[][]{
    \includegraphics[width=8.5cm]{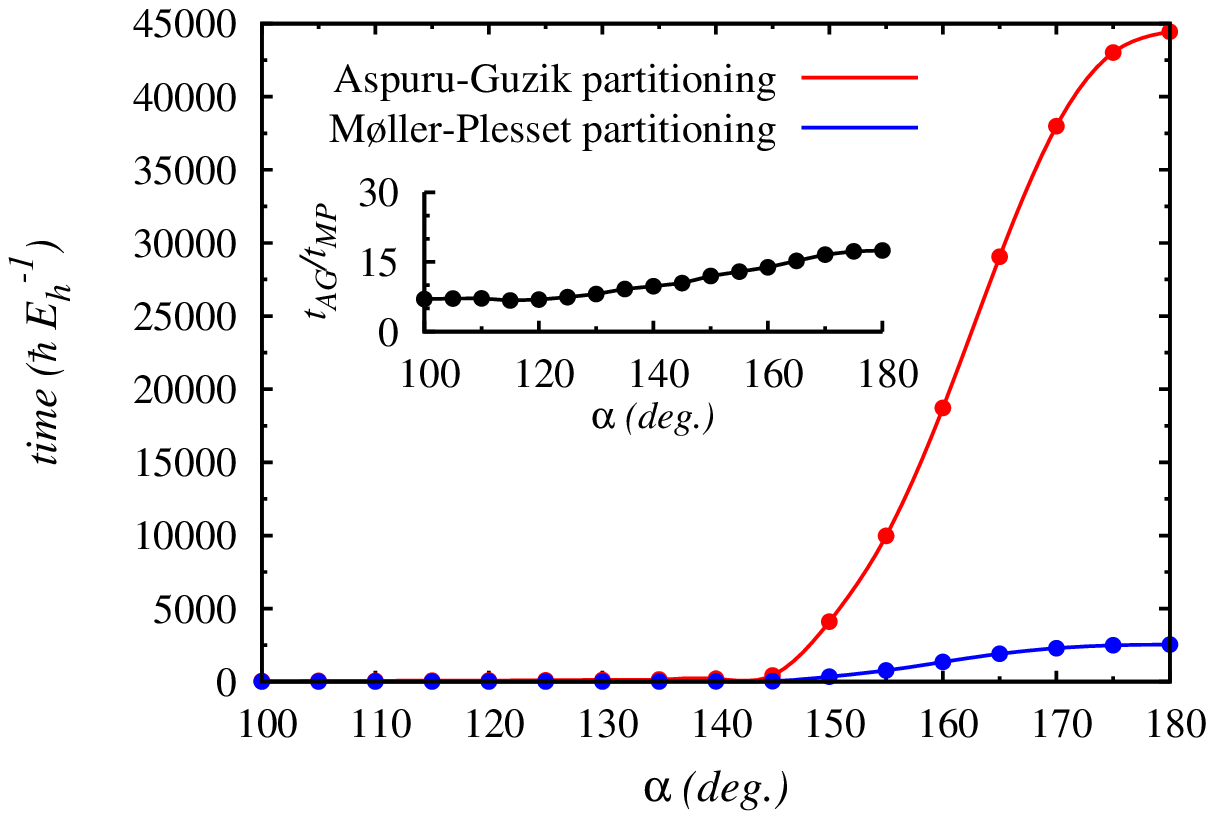}
    \label{time_bending}
  }
  \caption{Total ASP times (in atomic units) corresponding to 0.99 squared overlaps between the adiabatically prepared wave functions and the exact wave functions (for definition, see the text). (a) C-H bond stretching, (b) H-C-H angle bending. Red lines represent the original ASP \cite{aspuru-guzik_2005}, blue lines correspond to the M{\o}ller-Plesset type of initial Hamiltonians. The insets represent the ratio between these two ASP times.}
  \label{time_graphs}
\end{figure*}

The total times of ASP leading to the 99 \% squared overlap between the prepared and the exact wave functions are summarized in Figure \ref{time_graphs}. Both types of initial Hamiltonians are compared (at the level of theory mentioned above). As can be seen in the figure, ASP with MP-type of initial Hamiltonians (\ref{h_mp}) is superior to the original ASP \cite{aspuru-guzik_2005} in all simulated cases. For the whole H-C-H angle bending process and a large part of the C-H bond stretching (up to $r/r_0 = 2.5$) the total ASP times are smaller by a factor higher than 10 when using the MP initial Hamiltonians. For more stretched C-H bonds, ASP with the MP initial Hamiltonians is roughly two times faster.

\begin{figure*}
  \subfloat[][]{
    \hskip -1cm
    \includegraphics[width=9cm]{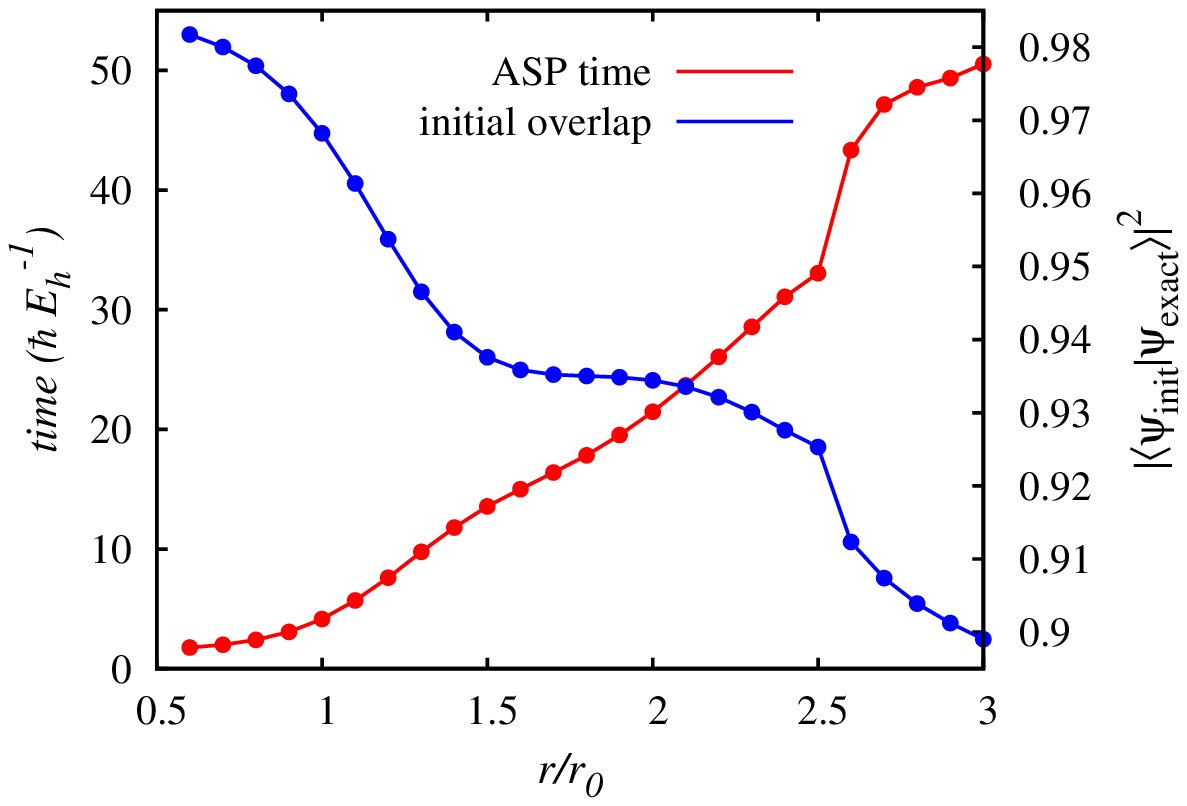}
    \label{time_disociace_cas66_init_guess}
  }
  \subfloat[][]{
    \includegraphics[width=9cm]{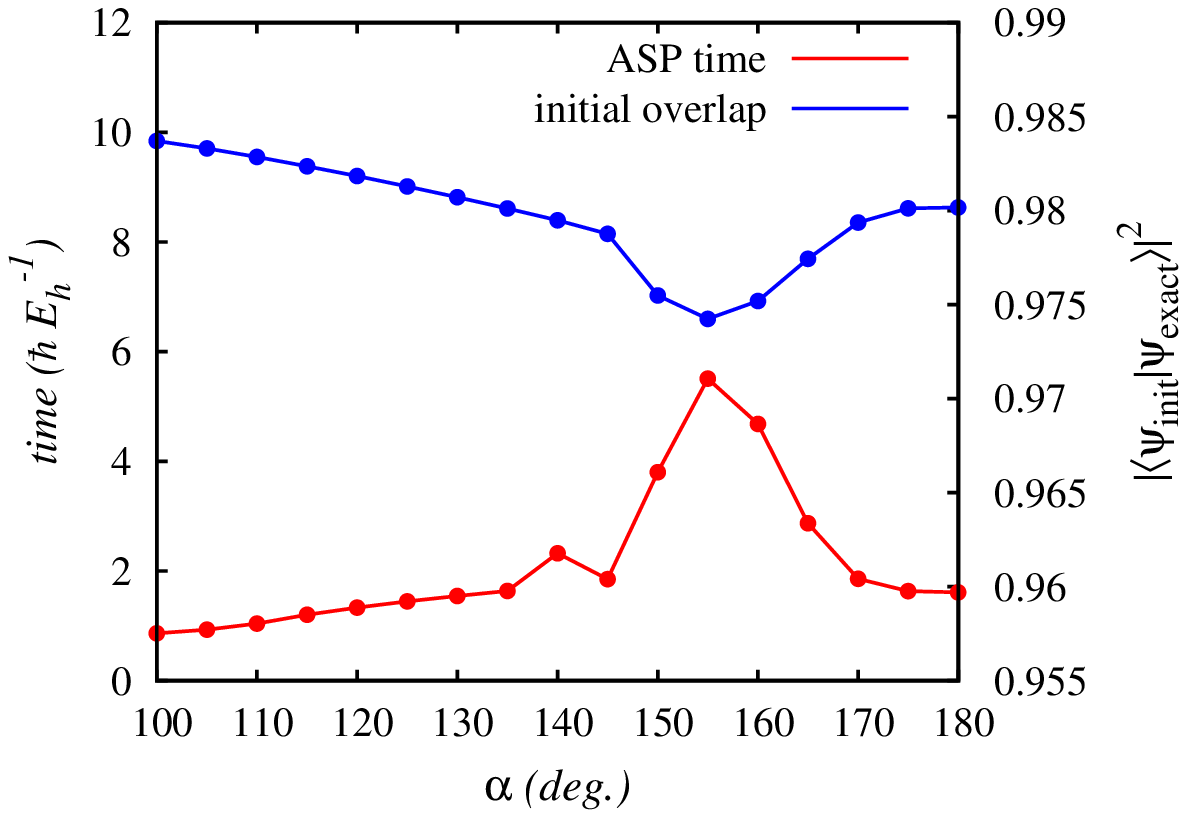}
    \label{time_bending_cas22_init_guess}
  }
  \caption{Total ASP times (red curves) corresponding to the 99 \% squared overlap between the prepared and the exact wave functions [CAS(6,12)/CAS(6,11)] for (a) C-H bond stretching with CASCI(6,6) initial guesses, and (b) H-C-H angle bending with CASCI(2,2) initial guesses. Blue curves represent the squared overlap between the initial and the exact wave function.}
  \label{time_cas_init_guess}
\end{figure*}

We have also studied ASP employing the CASCI initial wave functions with a limited CAS that cover the major part of a static correlation. These results are collected in Figure \ref{time_cas_init_guess}. In case of the H-C-H angle bending (Figure \ref{time_bending_cas22_init_guess}), the CAS(2,2) containing the quasi-degenerate HOMO and LUMO orbital pair has been used, while for the C-H bond stretching (Figure \ref{time_disociace_cas66_init_guess}) this orbital space has been augmented by bonding and anti-bonding orbitals of both C-H bonds resulting \mbox{in the CAS(6,6)}. One can see that when using these initial guesses, the total ASP time is reduced by four orders of magnitude for the linear geometry of CH$_2$ and by a factor of approximately $500$ in the worst case of the bond stretching. The squared overlaps between the initial and the exact [CAS(6,12)/CAS(6,11)] wave functions are also presented in Figure \ref{time_cas_init_guess}.

\begin{figure*}
  \subfloat[][]{
    \hskip -1cm
    \includegraphics[width=8.5cm]{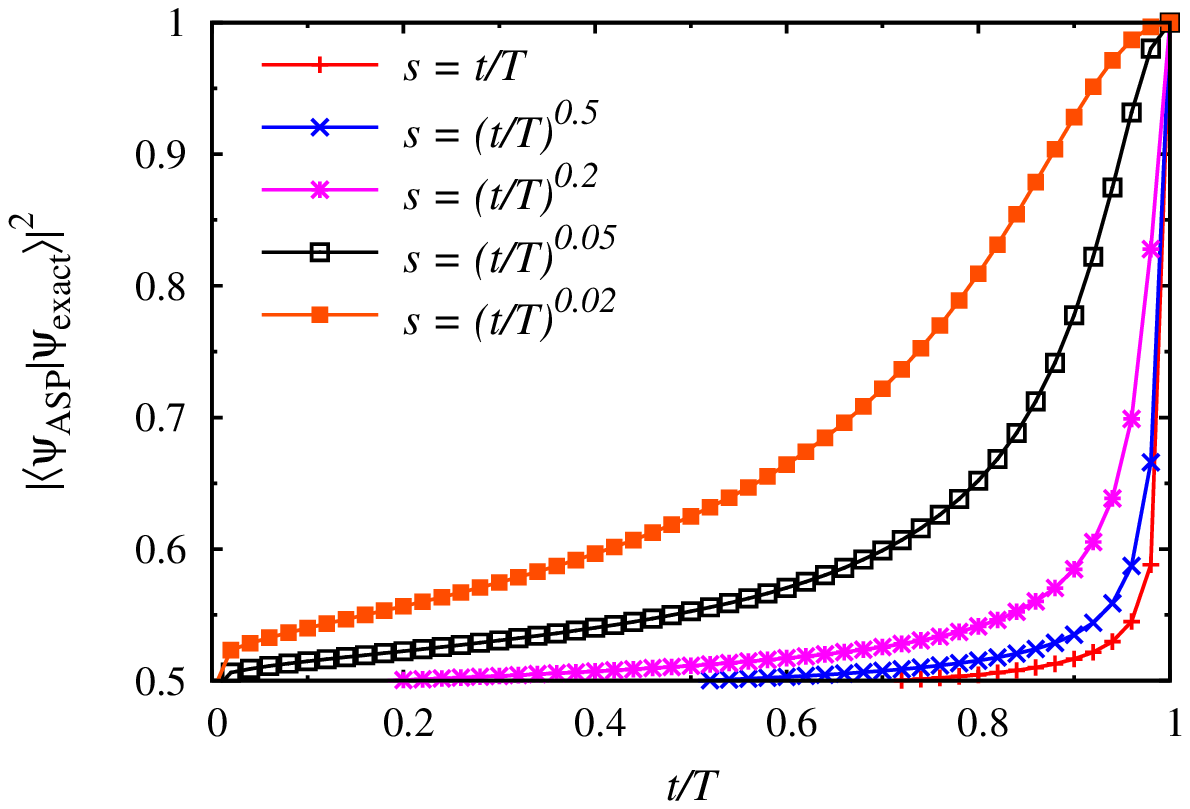}
    \label{nonlinear_diag}
  }
  \subfloat[][]{
    \includegraphics[width=8.5cm]{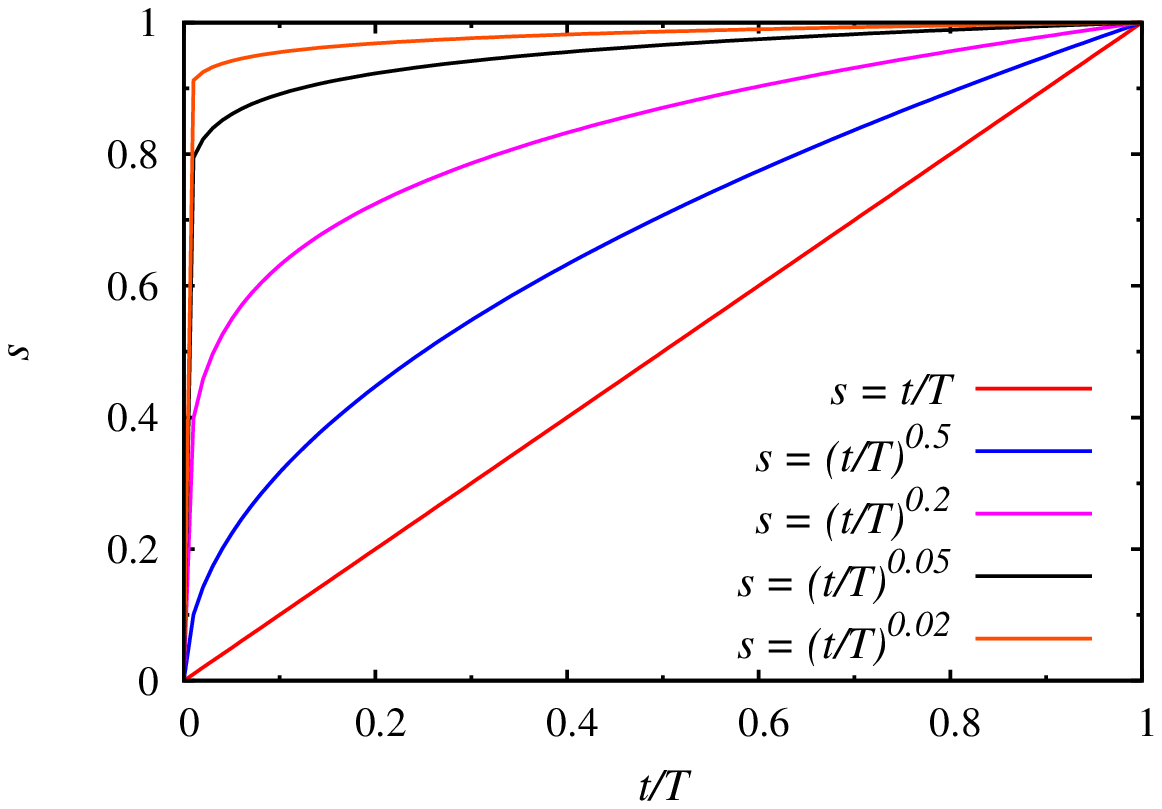}
    \label{nonlinear_fce}
  }
  \caption{(a) Squared overlap between the ASP \cite{aspuru-guzik_2005} wave function and the exact wave function [CAS(6,11)] for H-C-H angle bending with $\alpha = 180^{~\circ}$ and different nonlinear interpolations that are shown in (b).}
  \label{nonlinear}
\end{figure*}

The characteristic steep-at-the-end shape of the overlap between the adiabatically prepared and the exact wave functions as shown in Figures \ref{asp_graphs} and \ref{time_convergence} naturally leads one to the idea of a non-linear interpolation ASP path that would make this shape as gradual as possible. In other words, we may expect that rather than changing the Hamiltonian with a constant velocity, it would be advantageous to do a large part of ASP, where the overlap nearly does not change, quickly and slow down only at the problematic region close to $s = 1$. Figure \ref{nonlinear_fce} shows some of nonlinear interpolation paths with such an effect, whereas an impact on the shape of the overlap during ASP for the linear geometry of CH$_2$ is presented in Figure \ref{nonlinear_diag}.

\begin{figure}
  \includegraphics[width=9cm]{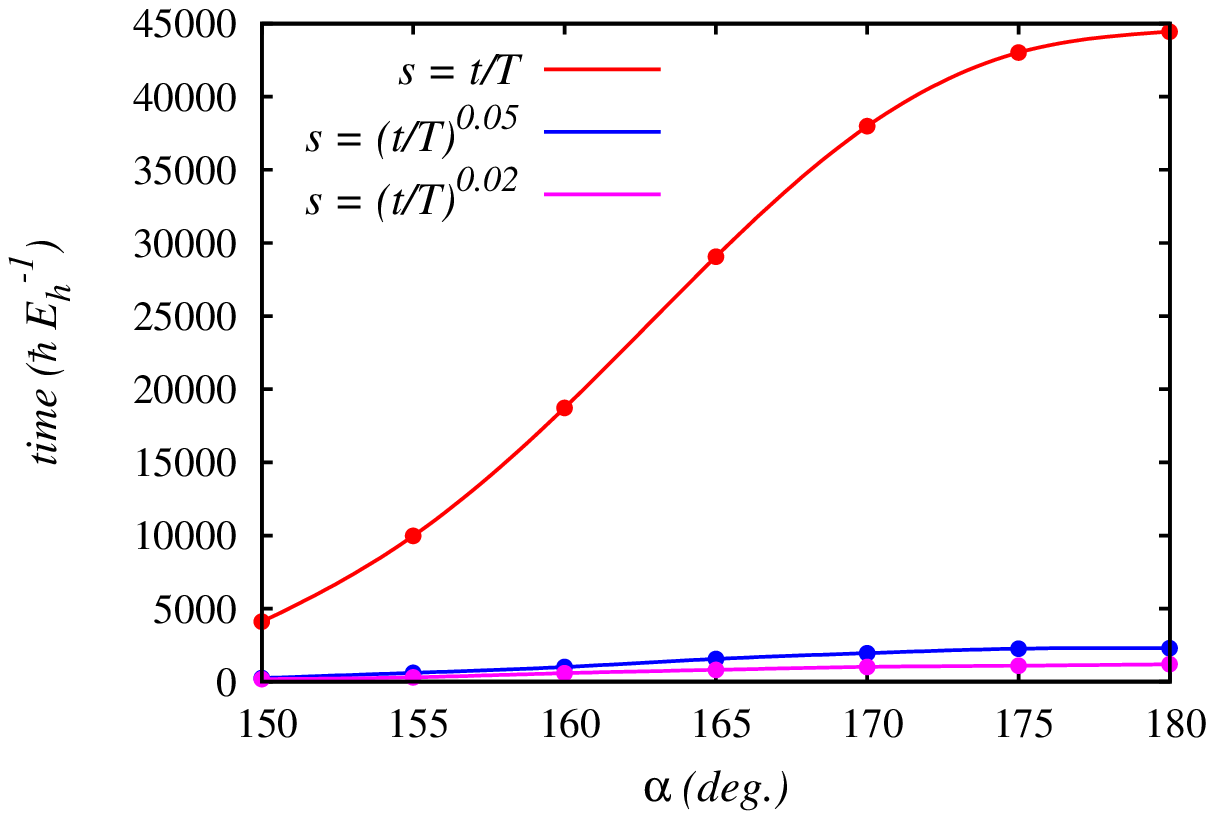}
  \caption{Comparison of the total ASP times (in atomic units) corresponding to 0.99 squared overlap between the adiabatically prepared and the exact [CAS(6,11)] wave functions for linear and nonlinear interpolation paths and H-C-H angle bending.}
  \label{time_bending_nonlinear}
\end{figure}

Figure \ref{time_bending_nonlinear} compares the total ASP times that correspond to the linear and two of the best nonlinear interpolation paths from Figure \ref{nonlinear}. Only the problematic part ($\alpha > 150^{~\circ}$), where the linear and nonlinear interpolations differ substantially, is shown. As can be seen, the non-linear approach with $s = (t/T)^{0.02}$ decreases the total ASP time at the linear geometry by a factor of approximately 40.

\begin{figure*}
  \subfloat[][]{
    \hskip -1cm
    \includegraphics[width=8.5cm]{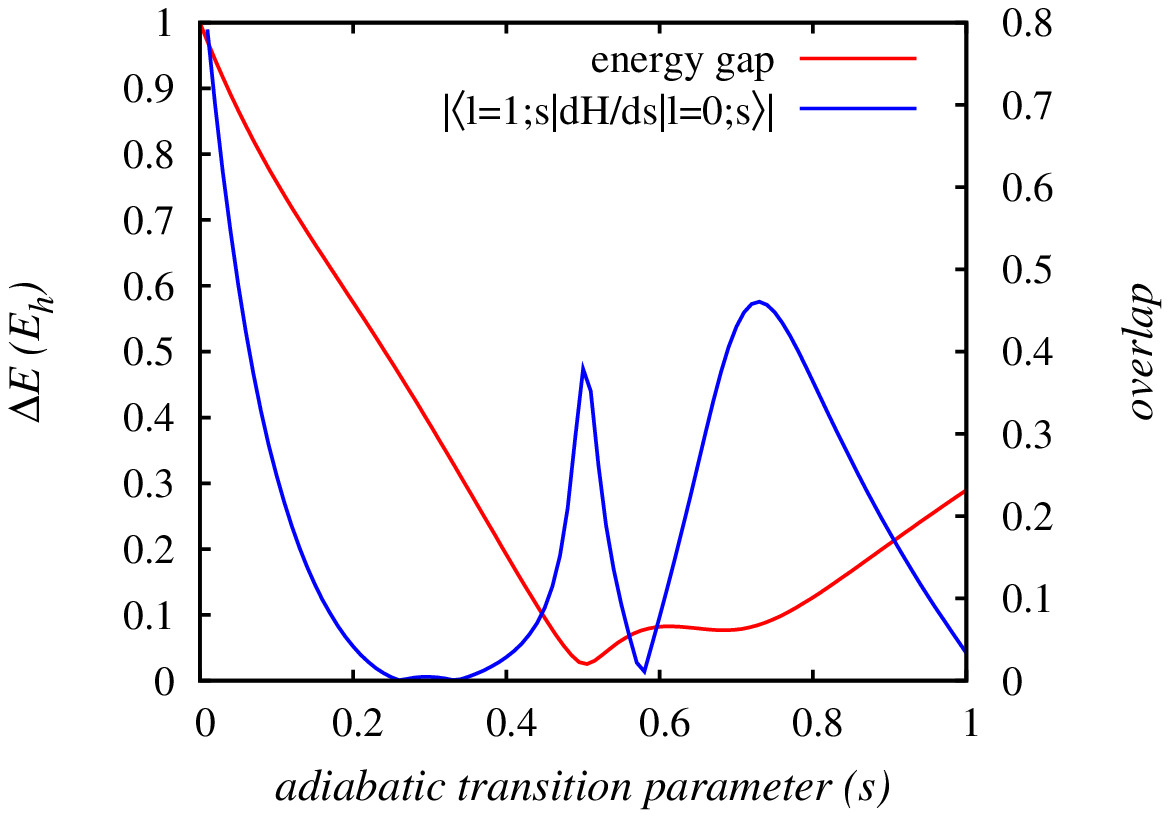}
    \label{xl_alpha_120}
  }
  \subfloat[][]{
    \includegraphics[width=8.5cm]{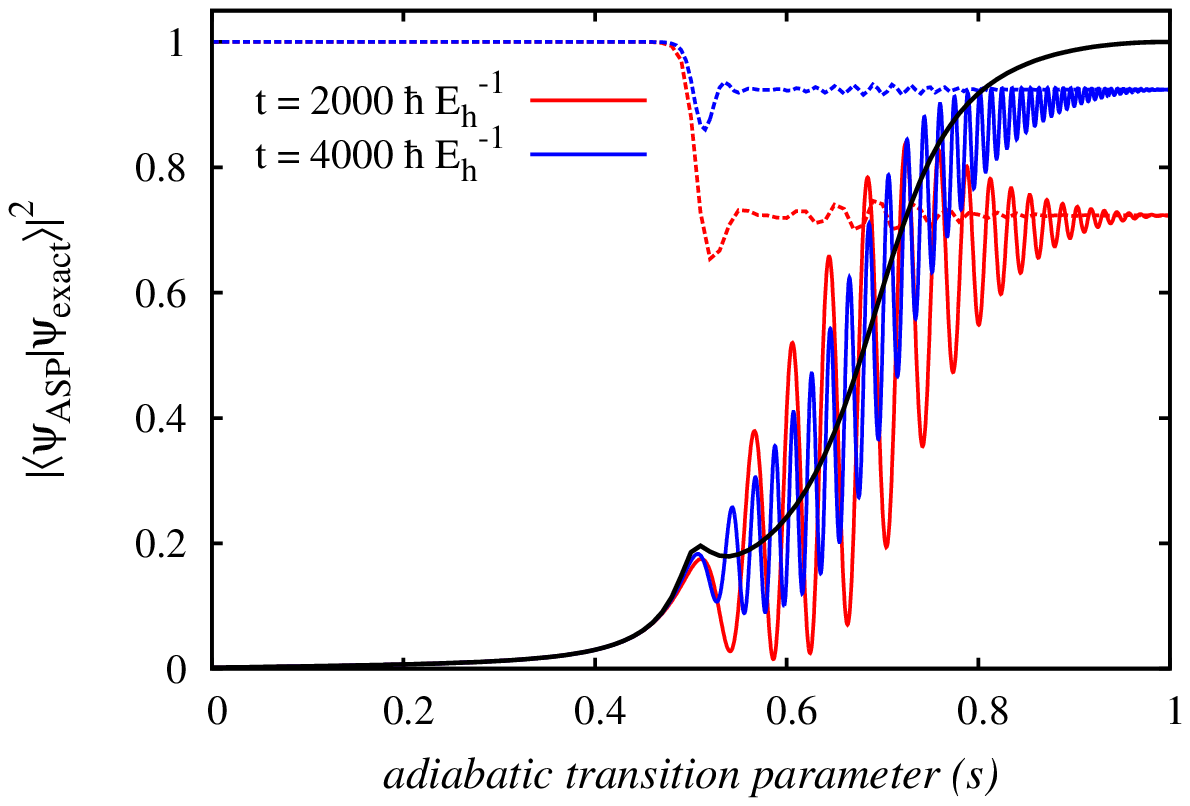}
    \label{xl_overlap_alpha_120}
  }
  \caption{Results of ASP with initial Hamiltonians composed of local $X$ fields corresponding to the sample geometry with $\alpha = 120~^{\circ}$. (a) Dependence of the minimum energy gap and  $|\bra{l=1; s} dH/ds \ket{l=0; s}|$ on the adiabatic transition parameter ($s$). (b) Dependence of the squared overlap between the ASP wave function and the exact wave function [CAS(6,7)] for two different lengths of ASP (time in atomic units). Black curve corresponds to $|\langle l = 0; s | \psi_{\text{exact}} \rangle|^2$ while dotted curves to $|\langle l = 0; s | \psi_{\text{ASP}} \rangle|^2$.}
  \label{xl}
\end{figure*}

\begin{figure}
  \includegraphics[width=9cm]{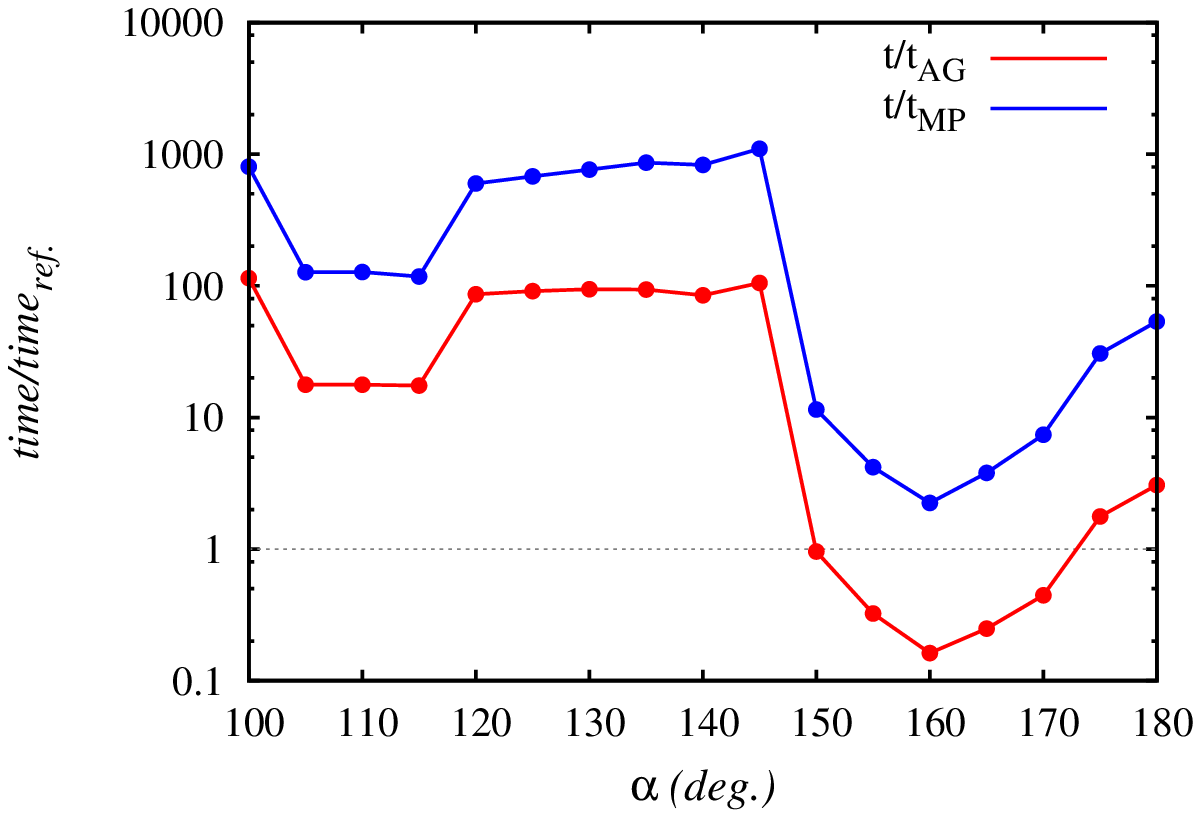}
  \caption{The ratio between total ASP times corresponding to initial Hamiltonians composed of local $X$ fields and times corresponding to the original ASP \cite{aspuru-guzik_2005} (red color) and M{\o}ller-Plesset type of initial Hamiltonians (blue color) for H-C-H angle bending.}
  \label{time_bending_lx}
\end{figure}

Figures \ref{xl} and \ref{time_bending_lx} summarize the results of ASP with initial Hamiltonians composed of local $X$ fields (\ref{hlx}). The progress of the minimum energy gap $g_{\text{min}}$ as well as the matrix element $\Big| \bra{l=1; s} \frac{d H}{ds} \ket{l=0; s} \Big|$ [its maximum value gives $\epsilon$ (\ref{epsilon})] during ASP are for the sample geometry with $\alpha = 120~^{\circ}$ depicted in Figure \ref{xl_alpha_120}. Figure \ref{xl_overlap_alpha_120} demonstrates for the same geometry the dependence of the squared overlap between the adiabatically prepared and exact wave functions for two different total times of ASP. The ratio between total ASP times (leading to the 99 \% squared overlap with the exact wave function) and times from Figure \ref{time_graphs} are shown in Figure \ref{hlx}.

\subsection{Discussion}

In the beginning of the previous section, we have numerically shown how the small overlap between the initial and the exact wave functions coming from a multireference character of a simulated state, together with a decreasing energy gap, complicate the ASP procedure. These properties manifest themselves in a very steep change of the overlap between the adiabatically prepared and the exact wave functions in the region close to $s = 1$. It is just this part of ASP that is difficult to follow adiabatically.

We have basically studied three possibilities of speeding up the original ASP procedure of Aspuru-Guzik et al. \cite{aspuru-guzik_2005} in these problematic regions, namely ($i$) taking initial Hamiltonians as a sum of the Fock operators (MP-type of initial Hamiltonians), ($ii$) using small-CAS-like initial wave functions (with proper initial Hamiltonians), ($iii$) and exploiting non-linear interpolation ASP paths. The speedup is always by a constant factor (independent of a system size) ranging from about 2 up to four orders of magnitude. 

\subsubsection{M{\o}ller-Plesset type of initial Hamiltonians}

A sufficient condition for the total ASP time (Eq. \ref{asp_time}) contains $\epsilon$ (\ref{epsilon}) in the numerator and minimal energy gap $g_{\text{min}}$ (\ref{gmin}) in the denominator. Since $\lVert \frac{d}{ds}H(s) \rVert$ is polynomial in a system size [as far as $H_{\text{final}}$ has an \textit{efficient} representation (\ref{ham_sec_quant})], critical is the dependence on $1/g^2_{\text{min}}$. It may in fact spoil the \textit{efficiency} of the algorithm when $g_{\text{min}}$ is exponentially small. However, this is what we do not expect for a typical molecular system \cite{aspuru_guzik_review}. 

In certain situations, the energy gaps can be amplified \cite{somma_2013}, but not in a general case considered here. Moreover, as Figure \ref{en_gap_graphs} indicates, minimal energy gaps of studied processes correspond to $s = 1$\footnote{The only exception is the most stretched geometry ($r/r_0 = 3.0$) where the minimum is not exactly at $s = 1$, but close to it. However, the energy difference between $\Delta E_{\text{min}}$ and $\Delta E(s = 1)$ is minor.} and thus cannot be avoided by a different ASP path. 

In \cite{aspuru-guzik_2005}, the authors used initial Hamiltonians with all matrix elements equal to zero except the only diagonal element equal to $E_{\text{HF}}$ with a reasoning that this approach yields initial gaps which are very large relative to typical electronic excitations. On the other hand, as the following simple analysis suggests, such an approach is not optimal for $\epsilon$. Due to the orthogonality of eigenvectors of $H(s)$, one needs $\frac{d}{ds}H(s)$ to be as close to identity as possible to minimize $\epsilon$. Since molecular Hamiltonians are diagonally dominated, $\frac{d}{ds}H(s)$ can approximate the identity matrix if all the diagonal elements change during ASP with a similar speed, or equivalently 

\begin{equation}
  \big[ (\mathbf{H}_{\text{final}})_{ii} - (\mathbf{H}_{\text{init}})_{ii} \big] \approx \big[ (\mathbf{H}_{\text{final}})_{jj} - (\mathbf{H}_{\text{init}})_{jj} \big], \quad \forall i,j. 
\end{equation}

\noindent
In the aforementioned approach \cite{aspuru-guzik_2005}, however, the diagonal element corresponding to the HF reference configuration does not change at all. 

On the contrary, when one uses initial Hamiltonian equal to a sum of the Fock operators (\ref{h_mp}) (initial wave function remains the same), the initial diagonal elements are equal to a sum of molecular orbital energies and all the diagonal elements during ASP change, which results in a smaller value of $\epsilon$. As is demonstrated in Figures \ref{asp_graphs} and \ref{time_graphs}, this approach makes the steep part of ASP more gradual and in average decreases the total ASP time by a factor of 10 (in case of the H-C-H angle bending and linear geometries by a factor of 20). The only region, where it is just two times faster, is that of more stretched C-H bonds. In this case, the HF reference configuration has a small contribution to the exact ground state wave function which means that both processes do not differ substantially. 

During the ASP method just described, one in fact adiabatically switches on the M{\o}ller-Plesset perturbation \cite{moller_plesset}. 

\subsubsection{Small-CAS-like initial wave functions}

Since in the most difficult potential energy surface (PES) regions corresponding to the H-C-H angle bending as well as the C-H bond stretching, the $\tilde{a}~^{1}A_{1}$ state of CH$_2$ exhibits very strong multireference character, it is a natural choice to employ other than HF initial wave functions that can be \textit{efficiently} precalculated on a classical computer and prepared on a quantum register, and cover as much of a static correlation as possible. When the number of strongly correlated electrons is rather small, as is the case of organic biradicals represented here by CH$_2$, the CASCI (or CASSCF) methods are adequate. When the number of strongly correlated electrons is larger (e.g. in transition metal compounds), the density matrix renormalization group (DMRG) method \cite{chan_review_2011} can be used. This is in analogy to classical multireference computational methods like CASPT2, where the dynamical correlation is calculated on top of the CASSCF wave function. From this perspective, such an approach can be viewed as an adiabatic inclusion of a dynamical correlation.

As can be seen in Figure \ref{time_cas_init_guess}, the speedups of up to four orders of magnitude (the case of H-C-H angle bending and linear geometries) can be achieved when employing small CASCI [CASCI(2,2)] initial wave functions that cover the major part of a static correlation. The explanation is simple: such initial wave functions correctly include static correlation and thus have much higher overlaps with the exact wave functions (dynamical correlation energy contribution is usually only about 1 \% of absolute energy, although its inclusion is essential for chemical accuracy) and the ASP path is therefore much shorter. To cover the major part of a static correlation in case of the C-H bond stretching, the CAS(6,6) composed of the HOMO, LUMO and two pairs of $\sigma$ bonding and anti-bonding  molecular orbitals (for both breaking C-H bonds) has to be used. As the initial overlap is smaller here (about 0.9 for the worst case), the speedup is also smaller and corresponds to a factor of 500 for the most stretched bonds. 

We have not dealt with particular forms of initial Hamiltonians here, we just note that as the active space contain small bounded number of spin orbitals, they can be implemented \textit{efficiently}.

\subsubsection{Non-linear interpolation paths}

The specific shapes of the squared overlap between the adiabatically prepared and the exact wave functions in problematic regions (see Figure \ref{asp_graphs}) lead us to use the non-linear interpolation paths that reserve most of the ASP time to the interval close to $s = 1$. As is demonstrated in Figure \ref{time_bending_nonlinear}, we were able to decrease the total ASP time approximately by a factor of 40 for the linear geometry of CH$_2$. 

However, we have to emphasize that this is not a general approach as it relies on the specific knowledge of how the ground state changes during ASP. Some sort of a systematic study would be desirable to find out whether the trends from Figure \ref{asp_graphs} are shared also by other multireference systems.

\subsubsection{Local X fields initial Hamiltonians}

The results of the original ASP \cite{aspuru-guzik_2005} as well as ASP with MP-type of initial Hamiltonians are consistent, locating minimum energy gaps and consequently the "steep" part that is difficult to follow adiabatically at $s = 1$. 

A speculative reason for this behavior might be the following. The MP-type of initial Hamiltonians in fact corresponds to the independent-particle model. In other words, the configuration interaction singles (CIS)
is the exact wave function of the excited state of the initial Hamiltonian. In quantum chemistry, it is known that CIS overestimates the excitation energy for correlated Hamiltonians, at least when the ground state is well described by a single Slater determinant. In case of the original ASP \cite{aspuru-guzik_2005}, the energy gap is maximum at the beginning by construction. It does not itself guarantee that the progress of the energy gap will be monotonous, but
one generally expects that when increasing the electron-electron correlation, the energy gap should decrease. 

The situation, however, changes dramatically when using initial Hamiltonians composed of local $X$ fields (\ref{hlx}). As is demonstrated in Figure \ref{xl_alpha_120} on the example of CH$_2$ geometry with $\alpha = 120~^{\circ}$, minimum energy gaps no longer occur at $s = 1$, but e.g. in this case close to $s = 0.5$. Moreover, the matrix element $\Big| \bra{l=1; s} \frac{d H}{ds} \ket{l=0; s} \Big|$ is rather high around $s = 0.5$, making together this region the most difficult part for ASP. Too fast change of a Hamiltonian (\ref{ham}) around $s = 0.5$ causes transitions to the first excited state $\ket{l=1;s}$, which is demonstrated in Figure \ref{xl_overlap_alpha_120} by decreasing of the overlap between the ASP wave function and the actual ground state $\ket{l=0;s}$. Figure \ref{xl_overlap_alpha_120} also shows the oscillatory character of the ASP wave function when expressed in the basis of eigenvectors of the exact Hamiltonian which is caused by the aforementioned transitions and decreases with increasing total ASP time.

The fact that minimum energy gaps does not occur at the end of the ASP procedure in fact complicates practical use of such an approach. As is depicted in Figure \ref{time_bending_lx} the dependence of total ASP times on $\alpha$ does not show any regular pattern as in Figure \ref{time_graphs}. Total times are for most cases much higher than total times of ASP with Hartree-Fock initial wave functions (up to three orders of magnitude). Only at the region between $\alpha = 150~^{\circ}$ and $\alpha = 170~^{\circ}$, total ASP times are accidentally slightly smaller than those of the original ASP procedure \cite{aspuru-guzik_2005} (but still higher than ASP with MP-type of initial Hamiltonians). We would like to note that situation is very the same also for the C-H bond stretching process, which is not displayed here. The only difference is that total ASP times of local X fields intital Hamiltonians are the highest in all cases.

In conclusion, our results indicate that quantum chemical initial states (and corresponding initial Hamiltonians) are important for ASP as they assure minimum energy gaps to occur at the end of the process. In other words, they make ASP time scaling with the HOMO-LUMO energy gap. Even Hartree-Fock single Slater determinants that have the overlap with the exact wave function smaller than 0.5 turns out to be sufficient (exceptions are the most stretched geometries with overlap around 0.2). These observations are true at least for the methylene, but as was mentioned above, some sort of a systematic multireference study should be done before making the statements general.

\section{Analysis when using two-body qubit Hamiltonians}
\label{experimental_proposal}
Here we analyze ASP of $\tilde{a}~^{1}A_{1}$ state of CH$_{2}$ in the CASCI(2,2) space when using two-body qubit Hamiltonians. Alternatively, as the number of active electrons as well as the number of active orbitals (and their symmetry properties) equal those of the hydrogen molecule in a minimal basis, the analysis is also valid for ASP of the FCI ground state of H$_2$ in a minimal basis. 

Unlike the recent proof-of-principle experimental realizations \cite{lanyon_2010,du_2010}, our proposal employs the direct mapping approach and can therefore, in contrast to the mentioned experiments which used explicit forms of Hamiltonian matrices, be simply adapted to other systems. We use the Jordan-Wigner transformation \cite{jordan_1928}, but have to emphasize that this is not an \textit{efficiently} scalable approach.
It turns out that for the \textit{efficient} direct implementation of the ASP algorithm, the Bravyi-Kitaev transformation, which balances locality of occupation and parity information \cite{seeley_2012}, is essential. When finishing this paper, we have learned about the very recent paper \cite{babbush_2013}, which in detail describes the \textit{efficient} algorithm relying on the Bravyi-Kitaev transformation. We will therefore not discuss why use of the Jordan-Wigner transformation is not \textit{efficient}, but rather refer the reader to \cite{babbush_2013}.

Details of the analysis can be found in Appendix \ref{experiment}. After the application of the Jordan-Wigner transformation, the Hamiltonian (\ref{ham_sec_quant}) of our model example consists of four single-qubit operations, six 2-qubit operations, and four 4-qubit operations [see (\ref{hs})]. We have used the perturbative gadgets technique \cite{jordan_2008} to transform the 4-qubit terms to 2-qubit terms at the cost of 16 ancilla qubits. For a detailed comparison of different types of gadgets with the emphasis on an experimental accessibility, see \cite{cao_2013}. Our proposal thus requires a total number of 20 qubits. 

\begin{figure}
  \includegraphics[width=8.5cm]{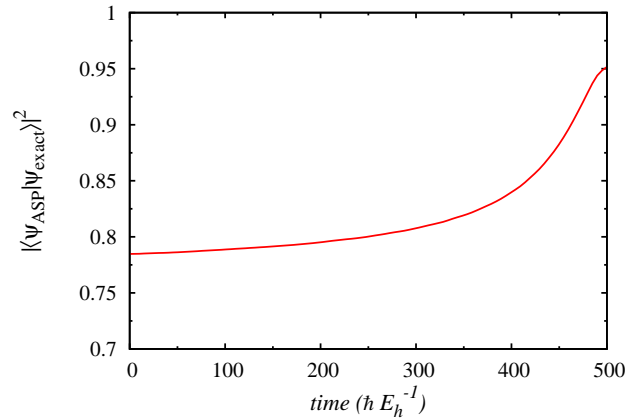}
  \caption{Numerical simulation of the proposed model ASP of the lowest lying singlet state of CH$_2$ employing the direct mapping and requiring 12 qubits, which corresponds to $r/r_0 = 1$, $\alpha = 160~^{\circ}$, and $T = 500~\hbar E_h^{-1}$.}
  \label{graph_experiment}
\end{figure}

We have numerically simulated proposed ASP for the geometry corresponding to $r/r_0 = 1$ and $\alpha = 160~^{\circ}$. From computational reasons only two of the four 4-qubit terms were considered (see Appendix \ref{experiment}), which leads to 8 ancilla qubits and the total number of 12 qubits. The results are presented in Figure \ref{graph_experiment}. 

\section{Conclusions}

In this paper, we have numerically investigated ASP of $\tilde{a}~^{1}A_{1}$ state of methylene (CH$_2$), which is a prototype of a molecular system with a difficult biradical electronic structure typical for such species as transition states of chemical reactions. 

We have presented three possibilities of speeding up the original ASP \cite{aspuru-guzik_2005} of $\tilde{a}~^{1}A_{1}$ state of CH$_2$, namely use of the MP-type of initial Hamiltonians, use of small CAS-like initial wave functions covering the major part of a static correlation, and also exploiting non-linear interpolation ASP paths. Most importantly, with the CASCI initial wave functions we were able to achieve speedups of up to four orders of magnitude. We have to emphasize that the speedup is always by a constant factor and does not influences the \textit{efficiency} of the ASP algorithm, which critically depends on the lowest energy gap between the ground and the first excited states.

Finally, we have analyzed the ASP of $\tilde{a}~^{1}A_{1}$ state of CH$_2$ in the CASCI(2,2) space employing the direct mapping. With the help of perturbative gadgets \cite{jordan_2008}, we have transformed the ASP Hamiltonian to contain at most 2-qubit interactions. The total number of required qubits in this case equals 20. 

\section*{Acknowledgement}
This work has been supported by the Grant Agency of the Czech Republic - GA\v{C}R (203/08/0626).

\bibliographystyle{h-physrev}
\bibliography{kvantove_pocitace,clanek_ch2,cc,dmrg}

\appendix

\clearpage
\section{Decomposition of ASP Hamiltonian to two-qubit terms}
\label{experiment}
Our proposal concerns primarily ASP of $\tilde{a}~^{1}A_{1}$ state of CH$_{2}$ in the CASCI(2,2) space. When we denote the HOMO with molecular spin orbital (SO) indices 1 and 2, and the LUMO with SO indices 3 and 4, the Hamiltonian (\ref{ham_sec_quant}) can be expressed as (only the non-zero terms are shown)

\begin{eqnarray}
  H & = & h_{11} a_1^{\dagger}a_1 + h_{22} a_2^{\dagger}a_2 + h_{33} a_3^{\dagger}a_3 + h_{44} a_4^{\dagger}a_4
+ \nonumber \\
   & + & h_{1221} a_1^{\dagger}a_2^{\dagger}a_2a_1 + h_{3443} a_3^{\dagger}a_4^{\dagger}a_4a_3 + h_{1441} a_1^{\dagger}a_4^{\dagger}a_4a_1 + \nonumber \\
   & + & h_{2332} a_2^{\dagger}a_3^{\dagger}a_3a_2 + \nonumber \\
   & + & (h_{1331} - h_{1313}) a_1^{\dagger}a_3^{\dagger}a_3a_1 + (h_{2442} - h_{2424}) a_2^{\dagger}a_4^{\dagger}a_4a_2 + \nonumber \\
   & + & h_{1243} (a_1^{\dagger}a_2^{\dagger}a_4a_3 + a_3^{\dagger}a_4^{\dagger}a_2a_1) + \nonumber \\ 
   & + & h_{1423} (a_1^{\dagger}a_4^{\dagger}a_2a_3 + a_3^{\dagger}a_2^{\dagger}a_4a_1),
  \label{ch2_ham}
\end{eqnarray}

\noindent
where

\begin{equation}
  h_{pqrs} \equiv \langle pq | sr \rangle = \int \mathrm{d}\mathbf{x}_1 \mathrm{d}\mathbf{x}_2 \chi_{p}^{*} (\mathbf{x}_1) \chi_{q}^{*} (\mathbf{x}_2) \frac{1}{r_{12}} \chi_{s} (\mathbf{x}_1) \chi_{r} (\mathbf{x}_2),
\end{equation}

\noindent
$\chi$ represent molecular spin orbitals and the values of individual integrals corresponding to the geometry with $r/r_0 = 1$ and $\alpha = 160^{~\circ}$  are summarized in Table \ref{integrals}. As the Hamiltonian (\ref{ch2_ham}) is shared also by the simplest molecular FCI system, the hydrogen molecule in a minimal basis \cite{whitfield_2010}, our proposal is valid for ASP of the ground state of H$_2$ as well. Table \ref{integrals} contains also the integral values for the hydrogen molecule adopted from \cite{whitfield_2010}. 

\begin{table}
  \begin{tabular}{c c c}
  \hline
  \hline
  Integrals & Value (CH$_2$) & Value (H$_2$) \\
  \hline
  $h_{11} = h_{22}$ & -0.853007 & -1.252477 \\[0.1cm]
  $h_{33} = h_{44}$ & -0.841410 & -0.475934 \\[0.1cm]
  $h_{1221}$ & 0.530171 & 0.674493 \\[0.1cm]
  $h_{3443}$ & 0.529723 & 0.697397 \\[0.1cm]
  $h_{1331} = h_{1441} = h_{2332} = h_{2442}$ & 0.481270 & 0.663472 \\[0.1cm]
  $h_{1313} = h_{2424} = h_{1243} = h_{1423}$ & 0.032834 & 0.181287 \\
  \hline
  \hline
  \end{tabular}
  \caption{One and two-electron molecular (spin) orbital integrals for the CH$_2$ proposal and also its H$_2$ equivalent. Integral values are expressed in atomic units ($E_h$), the CH$_2$ example corresponds to $r/r_0 = 1$, $\alpha = 160^{~\circ}$ and H$_2$ to $r = 1.401~a_0$ \cite{whitfield_2010}.}
  \label{integrals}
\end{table}

The initial Hamiltonian (\ref{h_mp}) on the other hand reads

\begin{eqnarray}
  H_{\text{init,MP}} & = & \big( h_{11} + h_{1221} \big) a_1^{\dagger}a_1 + \big( h_{22} + h_{1221} \big) a_2^{\dagger}a_2 + \nonumber \\
   & + & \big( h_{33} + h_{1331} + h_{2332} - h_{1313} \big) a_3^{\dagger}a_3 + \nonumber \\ 
   & + & \big( h_{44} + h_{1441} + h_{2442} - h_{2424} \big) a_4^{\dagger}a_4
\end{eqnarray}

After the application of the Jordan-Wigner transformation \cite{jordan_1928}

\begin{equation}
 a_{n}^{\dagger} = \Bigg( \bigotimes_{j=1}^{n-1} \sigma_{z}^{j} \Bigg) \otimes \sigma_{-}^{n}, \quad a_{n} = \Bigg( \bigotimes_{j=1}^{n-1} \sigma_{z}^{j} \Bigg) \otimes \sigma_{+}^{n},
\end{equation}

\noindent
where $\sigma_{\pm} = 1/2(\sigma_{x} \pm i \sigma_{y})$ and the superscript denotes the qubit on which the matrix operates, and some simple algebraic manipulations (for more details, see \cite{whitfield_2010}), the ASP Hamiltonian of our model system can be rewritten in terms of Pauli $\sigma$ matrices

\begin{eqnarray}
  H_{\text{ASP}}(s) & = & H^{0,1,2}(s) + H^{4}(s), \label{hs} \\ \nonumber \\
  H^{0,1,2}(s) & = & c_1(s) I + c_2(s) \big( \sigma_z^1 + \sigma_z^2 \big) + c_3(s) \big( \sigma_z^3 + \sigma_z^4 \big) + \nonumber \\
   & + & c_4(s) \sigma_z^2 \sigma_z^1 + c_5(s) \big( \sigma_z^3 \sigma_z^1 + \sigma_z^4 \sigma_z^2 \big) + \nonumber \\ 
   & + & c_6(s) \big( \sigma_z^4 \sigma_z^1 + \sigma_z^3 \sigma_z^2 \big) + c_7(s) \sigma_z^4 \sigma_z^3, \\ \nonumber \\
  H^4(s) & = & c_8(s) \big( \sigma_x^4 \sigma_x^3 \sigma_y^2 \sigma_y^1 + \sigma_y^4 \sigma_y^3 \sigma_x^2 \sigma_x^1 \big) + \nonumber \\ 
   & + & c_9(s) \big( \sigma_x^4 \sigma_y^3 \sigma_y^2 \sigma_x^1 + \sigma_y^4 \sigma_x^3 \sigma_x^2 \sigma_y^1 \big),
  \label{h4}
\end{eqnarray}

\begin{equation}
  s: 0 \rightarrow 1, \nonumber
\end{equation}

\noindent
with

\begin{eqnarray}
  c_1(s) & = & h_{11} + h_{33} + \Big( 1 - \frac{3s}{4} \Big) h_{1221} + \frac{s}{4} h_{3443} + \nonumber \\ 
         & + & (2 - s) h_{1331} - \Big( 1 - \frac{s}{2} \Big) h_{1313} \nonumber \\
  c_2(s) & = & - \frac{h_{11}}{2} + \Big( 2 - \frac{5s}{2} \Big) h_{1331} + \Big( \frac{5s}{4} - 1 \Big) h_{1313} - \frac{s}{4} h_{1221} \nonumber \\
  c_3(s) & = & - \frac{h_{33}}{2} + \Big( 2 - \frac{5s}{2} \Big) h_{1331} + \Big( \frac{5s}{4} - 1 \Big) h_{1313} - \frac{s}{4} h_{3443} \nonumber \\
  c_4(s) & = & \frac{s \cdot h_{1221}}{4} \nonumber \\
  c_5(s) & = & \frac{s}{4} \big( h_{1331} - h_{1313}\big) \nonumber \\
  c_6(s) & = & \frac{s \cdot h_{1331}}{4} \nonumber \\
  c_7(s) & = & \frac{s \cdot h_{3443}}{4} \nonumber \\
  c_8(s) & = &  - \frac{s \cdot h_{1313}}{4} \nonumber \\
  c_9(s) & = & \frac{s \cdot h_{1313}}{4}. 
\end{eqnarray}

All the terms in (\ref{hs}) are constant, single, or 2-qubit, except those of $H^{4}$ (\ref{h4}) which are 4-qubit. As a first step towards an experimental realization, we have transformed the 4-qubit terms to 2-qubit. Other possibility would be to simulate ASP on a digital quantum computer. In such a case, there is no need for this transformation. Nevertheless, as the simulated noise-free time propagation of the hydrogen molecule in a minimal basis corresponding to $U=\text{exp}(-i H t)$ with $t=1$ a.u. requires, due to the Trotter approximation, hundreds of quantum gates \cite{whitfield_2010, seeley_2012}, simulated longer-time evolution with the time-dependent Hamiltonian $H_{\text{ASP}}$ would certainly require thousands or more of them and undoubtedly also some sort of quantum error correction (QEC) \cite{gaitan_book} that will further significantly increase this number. In spite of a very promising progress in the ion-trap digital quantum simulation \cite{lanyon_2011}, such requirements are still out of reach of the present-day quantum technology.

To transform the 4-qubit terms to 2-qubit, we have used the perturbative gadgets technique of Jordan and Farhi \cite{jordan_2008}. We will only sketch the main ideas and then show the final Hamiltonian. For a detailed description and derivations, we refer the reader to the original paper \cite{jordan_2008}.

With the perturbative gadgets of \cite{jordan_2008}, one increases the Hilbert space of the quantum register with ancilla qubits and then on this augmented space constructs the gadget Hamiltonian $H^{\text{gad}}$, which is composed only of 2-qubit interactions, and whose low energy spectrum mimics the spectrum of the original $k$-qubit Hamiltonian. From the construction of $H^{\text{gad}}$ that will shortly follow, it turns out that the original $k$-qubit interactions appear at $k^{\text{th}}$ order of perturbation theory.

A general $k$-qubit Hamiltonian on $n$ qubits can be expressed as a sum of $r$ terms,

\begin{equation}
  H^{\text{comp}} = \sum_{s=1}^r c_s H_s
\end{equation}

\noindent
with coefficients $c_s$ and $H_s$ coupling some set of $k$ qubits according to

\begin{equation}
  H_s = \sigma^{s,k} \ldots \sigma^{s,2} \sigma^{s,1}, 
\end{equation}

\noindent
where each operator $\sigma^{s,j}$ is of the form

\begin{equation}
  \sigma^{s,j} = \hat{n}_{s,j} \cdot \vec{\sigma}^{s,j},
\end{equation}

\noindent
where $\hat{n}_{s,j}$ is a unit vector and $\vec{\sigma}^{s,j}$ is a vector of Pauli $\sigma$ matrices operating on $j^{\text{th}}$ qubit in the set of $k$ qubits acted upon by $H_s$. In this general case, one has to introduce $k$ ancilla qubits for each $H_s$ term, thus $rk$ ancilla qubits in total.

We now for simplicity restrict ourselves to the only one of the 4-qubit terms from (\ref{hs}), e.g.

\begin{equation}
  H^{\text{comp}} = \sigma^k \ldots \sigma^2 \sigma^1 = \sigma_x^4 \sigma_x^3 \sigma_y^2 \sigma_y^1.  
  \label{hcomp}
\end{equation}

\noindent
In this case we have to introduce $k = 4$  ancilla qubits (their indices will be written in capital letters: $I,J,\ldots$). Formally, we can write

\begin{equation}
  H^{\text{comp}} = I^8 I^7 I^6 I^5 \sigma_x^4 \sigma_x^3 \sigma_y^2 \sigma_y^1.
\end{equation}

The gadget Hamiltonian is defined as \cite{jordan_2008}

\begin{equation}
  H^{\text{gad}} = H^{\text{anc}} + \lambda V,
\end{equation}

\noindent
where $H^{\text{anc}}$ acts on ancilla qubits

\begin{eqnarray}
  H^{\text{anc}} & = & \sum_{1 \le I < J \le k} \frac{1}{2} \big( I - \sigma_z^I \sigma_z^J \big) = \nonumber \\
   & = & 3I - \frac{1}{2} \Big( \sigma_z^{6} \sigma_z^{5} + \sigma_z^{7} \sigma_z^{5} + \sigma_z^{8} \sigma_z^{5} + \sigma_z^{7} \sigma_z^{6} + \nonumber \\
   & + & \sigma_z^{8} \sigma_z^{6} + \sigma_z^{8} \sigma_z^{7} \Big),
\end{eqnarray}

\noindent
and the perturbation $V$ couples ancilla and computational qubits

\begin{eqnarray}
  V & = & \sum_{j=1}^k \sigma_x^J \otimes \sigma^j = \nonumber \\
   & = & \sigma_x^5 \sigma_y^1 + \sigma_x^6 \sigma_y^2 + \sigma_x^7 \sigma_x^3 + \sigma_x^8 \sigma_x^4. 
\end{eqnarray}

\noindent
The perturbative expansion converges provided that

\begin{equation}
  \lambda < \frac{k - 1}{4k}.
\end{equation}

Since $H^{\text{gad}}$ commutes with $X = \sigma_x^8 \sigma_x^7 \sigma_x^6 \sigma_x^5$, it can be block diagonalized into blocks corresponding to $+1$ and $-1$ eigensubspaces of $X$. As is shown in \cite{jordan_2008} by means of a degenerate perturbation theory, when one uses the ancilla qubits in the state 

\begin{equation}
  \ket{+} = \frac{1}{\sqrt{2}} \big( \ket{0000} + \ket{1111} \big), 
  \label{anc_plus}
\end{equation}

\noindent
which corresponds to $+1$ eigensubspace of $X$, the low energy eigenstates of $H^{\text{gad}}$ approximate $H^{\text{comp}}$. In $+1$ eigensubspace of $X$, $H^{\text{anc}}$ has degeneracy $2^4$ and $\lambda V$ perturbs this ground space in two separate ways. Firstly, it shifts the energy of the entire space (which does not matter as we are interested in eigenstates, not the eigenvalues). Secondly, at $k^{\text{th}}$ order in perturbation theory, it splits the degeneracy. This splitting in fact allows the low energy subspace of $H^{\text{gad}}$ to mimic the spectrum of $H^{\text{comp}}$.

When all the terms of $H^4$ (\ref{h4}) are taken into account, 16 ancilla qubits (4 for each term) are necessary, thus 20 qubits in total. In such a case, each 4-qubit ancilla sub-register is initialized into the state (\ref{anc_plus}) and the state of ancilla qubits reads

\begin{equation}
  \ket{\text{ancilla}} = \ket{+} \otimes \ket{+} \otimes \ket{+} \otimes \ket{+}.
\end{equation}

\noindent
With a procedure analogous to that of just one 4-qubit term, $H^4(s)$ (\ref{h4}) can be expressed as

\begin{eqnarray}
  H^4(s) & = & k_s \cdot \Big[ 12 I -\frac{1}{2} \Big( \sigma_z^{6} \sigma_z^{5} + \sigma_z^{7} \sigma_z^{5} + \sigma_z^{8} \sigma_z^{5} + \nonumber \\
   & + & \sigma_z^{7} \sigma_z^{6} + \sigma_z^{8} \sigma_z^{6} + \sigma_z^{8} \sigma_z^{7} + \sigma_z^{10} \sigma_z^{9}  + \sigma_z^{11} \sigma_z^{9} + \nonumber \\ 
   & + & \sigma_z^{12} \sigma_z^{9} + \sigma_z^{11} \sigma_z^{10} + \sigma_z^{12} \sigma_z^{10} + \sigma_z^{12} \sigma_z^{11} + \nonumber \\
   & + & \sigma_z^{14} \sigma_z^{13} + \sigma_z^{15} \sigma_z^{13} + \sigma_z^{16} \sigma_z^{13} + \sigma_z^{15} \sigma_z^{14} \nonumber \\
   & + & \sigma_z^{16} \sigma_z^{14} + \sigma_z^{16} \sigma_z^{15}  + \sigma_z^{18} \sigma_z^{17} + \sigma_z^{19} \sigma_z^{17} + \nonumber \\
   & + & \sigma_z^{20} \sigma_z^{17} + \sigma_z^{19} \sigma_z^{18} + \sigma_z^{20} \sigma_z^{18} + \sigma_z^{20} \sigma_z^{19} \Big) + \nonumber \\
   & + & \lambda \Big( c_8(s) \sigma_x^{5} \sigma_y^{1} + \sigma_x^{6} \sigma_y^{2} + \sigma_x^{7} \sigma_x^{3} + \sigma_x^{8} \sigma_x^{4} + \nonumber \\ 
   & + & c_8(s) \sigma_x^{9} \sigma_x^{1} + \sigma_x^{10} \sigma_x^{2} + \sigma_x^{11} \sigma_y^{3} + \sigma_x^{12} \sigma_y^{4} + \nonumber \\
   & + & c_9(s) \sigma_x^{13} \sigma_x^{1} + \sigma_x^{14} \sigma_y^{2} + \sigma_x^{15} \sigma_y^{3} + \sigma_x^{16} \sigma_x^{4} + \nonumber \\ 
   & + & c_9(s) \sigma_x^{17} \sigma_y^{1} + \sigma_x^{18} \sigma_x^{2} + \sigma_x^{19} \sigma_x^{3} + \sigma_x^{20} \sigma_y^{4} \Big) \Big], \nonumber \\
\end{eqnarray}

\noindent
here $\lambda$ must satisfy

\begin{equation}
  \lambda < \frac{3}{64}.
\end{equation}

\noindent
and as the desired splitting appears at $k^{\text{th}}$ order in perturbation theory (the effect is rather weak), the gadget Hamiltonian has to be scaled by the factor 

\begin{equation}
  k_s = -\frac{6}{4 \lambda^4}, 
\end{equation}

\noindent
to approximate the eigenstates of $H_{\text{ASP}}(s)$ when added with $H^{0,1,2}(s)$.

\begin{figure}
  \includegraphics[width=9cm]{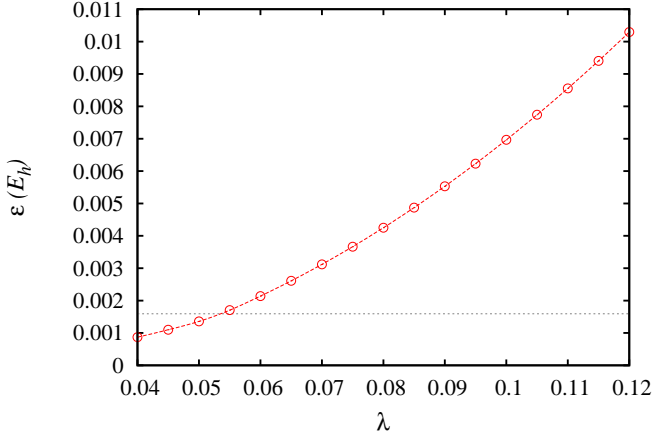}
  \caption{Dependence of the spectral error $\epsilon$ on the perturbation parameter $\lambda$ for the case of a numerical simulation shown in Figure \ref{graph_experiment} [$c_9(s) = 0$] and the ASP transition parameter $s = 1$. The horizontal dotted line corresponds to the accuracy of 1 kcal/mol.}
  \label{gadget_accuracy}
\end{figure}

In Section \ref{experimental_proposal}, we have presented the results of the numerical simulation, which for computational reasons implemented ASP with $H_{\text{ASP}}(s)$ containing only the first two 4-qubit terms [it is equivalent to setting $c_9(s) = 0$]. Such restriction corresponds to the total number of 12 qubits. The perturbation parameter $\lambda$ was in the numerical simulation set to $0.01$. The dependence of the spectral error $\epsilon$ (the maximum absolute deviation between the eigenspectrum of the gadget and target Hamiltonians) is for the simulated case and $s = 1$ presented in Figure \ref{gadget_accuracy}, showing that the value of $\lambda$ we have employed guarantees that ``chemical accuracy" is achieved.  

\end{document}